\documentclass[12pt]{article}
\usepackage{newinutile}
\usepackage{amsmath,amssymb}
\usepackage{epsfig,graphics,color,calc,graphicx}
\usepackage{cite}
\usepackage{mathrsfs}
\usepackage[title]{appendix}

\def\theequation{\arabic{section}.\arabic{equation}}

\newcommand{\be}{\begin{equation}}
\newcommand{\ee}{\end{equation}}
\newcommand{\bea}{\begin{eqnarray}}
\newcommand{\eea}{\end{eqnarray}}
\newcommand{\bpr}{\begin{proof}}
\newcommand{\epr}{\end{proof}}
\renewcommand{\Re}{\mathrm{Re}}

\begin{document}
\title{Transport in quantum multi--barrier systems as random walks on a lattice}

\author{Emilio N.\ M.\ Cirillo}
\email{emilio.cirillo@uniroma1.it}
\affiliation{Dipartimento di Scienze di Base e Applicate per l'Ingegneria, \\
             Sapienza Universit\`a di Roma, 
             via A.\ Scarpa 16, I--00161, Roma, Italy.}

\author{Matteo Colangeli}
\email{matteo.colangeli1@univaq.it}
\affiliation{Dipartimento di Ingegneria e Scienze dell'Informazione e 
Matematica,\\
 Universit\`a degli Studi dell'Aquila, 
via Vetoio, 67100 L'Aquila, Italy.}

\author{Lamberto Rondoni}
\email{lamberto.rondoni@polito.it}
\affiliation{Dipartimento di Scienze Matematiche, 
Dipartimento di Eccellenza 2018--2022,\\
Politecnico di Torino,
corso Duca degli Abruzzi 24, I--10129, Torino, Italy. \\ 
INFN, Sezione di Torino, Via P. Giuria 1, 10125 Torino, Italy.}

\begin{abstract}
A quantum finite multi--barrier system, {with a periodic potential}, is considered and exact 
expressions 
for its plane wave amplitudes 
are obtained using the Transfer Matrix method \cite{CNP15}.
This quantum model is then associated with a stochastic process of independent random walks on a lattice, by properly
{relating the wave amplitudes} with the hopping probabilities of the particles
moving on the lattice and with the injection rates from external particle reservoirs. Analytical and numerical results 
prove that the stationary density
profile of the particle system overlaps with the quantum mass density  
profile of the stationary Schr\"{o}dinger equation, 
when the parameters of the two
models are suitably matched. The equivalence between 
the quantum model and a stochastic particle system
would mainly be fruitful in a
disordered setup.
Indeed,
we also show, here, that this connection, analytically proven to hold
for periodic barriers, holds even when the width of the barriers and the 
distance between barriers are randomly chosen.
\end{abstract}



    

\maketitle

\vfill\eject

\section{Introduction}
\label{s:introduzione}
\par\noindent
A variation of the Anderson model \cite{anderson} for disordered 
solids has been recently introduced 
and investigated 
in order to describe systems that are finely structured, but not macroscopic 
\cite{col1,col2,VanRon}. In particular, for 1D chains of $N$ potential barriers, 
the large $N$ limit was performed in a way that prevents the application of standard techniques, 
such as Furstenberg type theorems \cite{CRP}. Numerically, it has been observed that such systems 
do not lead to localization, as indeed natural for systems whose length remains finite and fixed, 
when the number of barriers grows. Unlike the tight--binding model introduced by Anderson, the one 
of Refs.~\cite{col1,col2,VanRon} enjoys a purely off--diagonal disorder that affects the tunneling 
couplings among the wells, but not the energies of the bound states within the wells. 
In Refs.~\cite{col1,col2}, it has been numerically shown that the $N \to \infty$ limit for the model 
there treated leads to the conclusion that a large deviation principle holds for the fluctuations of
the transmission coefficient, with a proper scaling for the rate function. 
In Ref.~\cite{VanRon}, 
the transmission coefficient has been investigated for physically relevant numbers of potential 
barriers, referring to the scales of nano--structured sensors. In particular, the effect of compression
has been numerically investigated, for different compression rules, finding that the rate of convergence 
to the behavior of the Kronig--Penney model \cite{KP1931,CNP15} strongly depends on the disorder degree; 
that compression induces a decrease of the transmission coefficient; and that a moderate number of barriers  
together with strong disorder imply high sensitivity to compression, which can be used for pressure sensors.

In search for a more detailed treatment of the model of 
Refs.~\cite{col1,col2,VanRon}, the present paper 
investigates the possibility of adopting an effective description, 
that replaces the original one with
models that behave in an equivalent fashion, while being better suited for 
analytical treatment. The first 
case that ought to be investigated, in this respect, is the regular one, corresponding to the finite
Kronig--Penney model.
We show that the wave function time independent 
profile of the Kronig--Penney 
model can be associated with the stationary particle profile 
of independent random walkers with suitable hopping rates. 
For convenience, we deal with this model, using the language 
of the Zero Range Process (ZRP) \cite{EH2005,LS99,LMS04,CC17,CCM16,CCM16pre}. 
The regular quantum system considered here is the same as the one studied in Ref.\cite{CNP15}, 
with $N$ potential barriers and fixed length $L$. Here, we focus on the connection between 
both regular and disordered quantum models of that kind with properly devised stochastic processes. 
This connection is established by introducing site-dependent hopping 
probabilities for the stochastic models, in terms of the parameters of the
corresponding quantum systems. 
This connection allows the analysis of one of the models in terms of the other, which is 
particularly useful in the disordered cases.

The paper is organized as follows. In Section~\ref{s:mod}, we illustrate our model, which consists of 
a 1D sequence of $N$ regularly placed conducting and insulating regions, and we solve the 
time independent Schr\"odinger equation for the density of particles, at the center of each conducting
region, as a function, in particular, of $N$ and of the energy of the particles $E$. We 
observe that the transport of particles is approximately ballistic when the energy $E$ is sufficiently 
higher than a reference energy $E_0$.

In Section~\ref{s:zrp}, we 
summarize the fundamental properties of 
the Zero Range Process, and we show how the parameters of the 
quantum and of the stochastic process
can be tuned so that their profiles coincide. 
In Section~\ref{s:num}, we test numerically 
the validity of our method comparing the Monte Carlo results with 
the exact expressions. We show that our technique 
also applies to the case of random environments, 
for which no general ergodic--like results seem to have been so far developed.
In particular we show that, for sufficiently large $N$ and $L$ fixed, 
the quenched ZRP averaged profiles tend to stationary regular ones, implying, 
also, that in the equivalent quantum model no 
localization occurs.
This happens because our large $N$ limit is not of the standard hydrodynamic kind; rather, it is meant to describe small systems.
Conclusions are drawn in Section~\ref{s:concl}.

\section{The quantum multi--barrier model}
\label{s:mod}
\par
\begin{figure}[h]
\centering
\includegraphics[width = 0.7\textwidth]{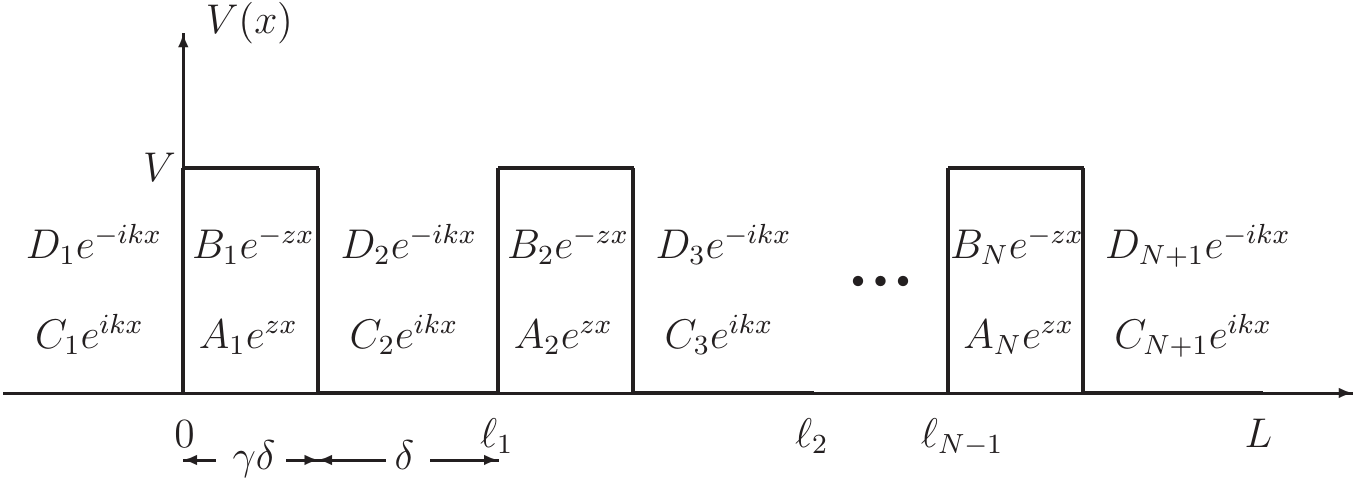} 
\caption{Schematic representation of the finite Kronig--Penney model.}
\label{fig:fig0}
\end{figure}

\noindent
Consider a 1D medium consisting of $N$ identical potential barriers of width $\gamma\delta$
separated by a distance $\delta$, 
cf.\ Fig.~\ref{fig:fig0}, \cite{CNP15}. The total length of the system is 
\begin{equation}
\label{elle}
L = (1+\gamma) \delta N 
\end{equation}
and the position of the left side of the $n$--th barrier is given by:
\begin{equation}
\label{ellen}
\ell_n= n \delta(1+\gamma) \quad \text{with} \quad n=0,\dots,N 
\;\;.
\end{equation}
For the potential barrier, we take $V(x)=V > 0$ for $x\in[\ell_{n-1},\ell_{n-1}+\gamma\delta]$, 
$n=1,\dots,N$, and $0$ for the other positions $x$. 
The zero potential regions
can be interpreted 
as electrically conducting regions and 
the potential barriers as made of a given insulating material. 
As in Refs.~\cite{col1,col2,VanRon,CNP15}, the total length $L$, and
the fraction  $\gamma$ of insulating to conducting material are fixed, 
while the number of barriers $N$ can be varied to 
represent more or less finely structured media. 
Therefore, the width of the conducting regions is given by
\begin{equation}
\label{delta}
\delta\equiv\delta_N=\frac{1}{1+\gamma}\frac{L}{N} 
\;\;.
\end{equation}
Like $\delta_N$, most of the quantities introduced in this paper depend 
on $N$, 
but for sake of notation simplicity we omit to 
explicitly note it where there is no
risk of confusion.

In Ref.~\cite{CNP15}, it was found that a single barrier of height 
\begin{equation}
E_0=\frac{\gamma}{1+\gamma}V
\label{zero}
\end{equation} 
leads to the same asymptotic transmission coefficient of 
the Kronig--Penney model
in the $\gamma,L$--\emph{continuum limit}, 
namely, when all the parameters of the model are kept fixed but the 
number of barriers $N$ which tends to infinity.
The potential $E_0$, which was there called
{\em zero point energy} in Ref.~\cite{CNP15}, will be used 
also here to discriminate various
transport regimes.

We now write the stationary 
Schr\"odinger equation in units such that the constant $\hbar^2/2 m$ 
($m$ the mass of the electron) equals 1, 
denoting by a prime the space derivatives:
\begin{equation}
\label{qua000}
-\psi''(x)+ V(x) \psi(x) = E\psi(x)
\;\;.
\end{equation}

As in Ref.~\cite{CNP15}, the solution corresponding to positive energy, $E>0$,
can be written as 
\begin{equation}
\label{qua010}
\psi(x)
=
\left\{
\begin{array}{ll}
C_1e^{ikx}+D_1e^{-ikx}
&
x\le0\\
A_ne^{zx}+B_ne^{-zx}
&
\ell_{n-1}<x<\ell_{n-1}+\gamma\delta\\
C_{n+1}e^{ikx}+D_{n+1}e^{-ikx}
&
\ell_{n-1}+\gamma\delta\le x\le \ell_n\\
C_{N+1}e^{ikx}+D_{N+1}e^{-ikx}
&
x> \ell_N\\
\end{array}
\right.
\end{equation}
where $n=1,\dots,N$, 
$k=\sqrt{E}$,
and $z=\sqrt{V-E}$.
The boundary conditions provide the real constants $C_1$ and $D_{N+1}$, below simply denoted by $C$ and $D$ 
respectively. The remaining coefficients $C_n$ and $D_n$, that respectively represent the amplitudes of 
the waves entering and being reflected from the left boundary of the 
$n$--th barrier, are then computed 
in terms of $C$ and $D$, imposing the continuity of both the wave function and its first derivative.

As in \cite{CNP15}, we introduce the notation 
\begin{equation}
\mathbf{\Delta}(a)=\left( \begin{array}{c c}
e^{a} & 0 \\
0 &  e^{-a}
\end{array} \right) ~, \quad \forall a\in\mathbb{C}
\label{qua020}
\end{equation} 
and we observe that 
\begin{equation}
\label{qua030}
\mathbf{\Delta}(a)^{-1}=\mathbf{\Delta}(-a),\;\;
\mathbf{\Delta}(a)\mathbf{\Delta}(b)
=
\mathbf{\Delta}(a+b),\;\;
\mathbf{\Delta}(0)=\mathbb{I},\;\;
\det(\mathbf{\Delta}(a))=1 ~, \quad \forall a,b \in \mathbb{C}
\end{equation}
where $\mathbb{I}$ denotes the identity.
Moreover, we set
\begin{equation}
\label{qua040}
\mathbf{T}(a,b)
= \left(\begin{array}{cc}e^{ab}&e^{-ab}\\ae^{ab}&-ae^{-ab}\\\end{array}\right) ~, \quad \forall a,b \in \mathbb{C}
\end{equation}
and we remark that 
for any $a,b,c\in\mathbb{C}$
\begin{equation}
\label{qua050}
\mathbf{T}(a,b)
=
\mathbf{T}(a,0)\mathbf{\Delta}(ab),\;
\mathbf{T}(a,b+c)
=
\mathbf{T}(a,b)\mathbf{\Delta}(ac),\;
\det(\mathbf{T}(a,b))=-2a,\;
\end{equation}
and
\begin{equation}
\label{qua055}
\mathbf{T}(a,0)^{-1}=\mathbf{T}(1/a,0)^\dag/2
\end{equation}
where $\dag$ denotes matrix transposition.

With such notation, 
the continuity of the wave function and its first derivative at 
the points $\ell_n$ and $\ell_n+\gamma\delta$ can be written as 
\begin{displaymath}
\mathbf{T}(ik,\ell_n)
\left(\begin{array}{c}C_{n+1}\\D_{n+1}\\\end{array}\right)
=
\mathbf{T}(z,\ell_n)
\left(\begin{array}{c}A_{n+1}\\B_{n+1}\\\end{array}\right)
\end{displaymath}
and
\begin{displaymath}
\mathbf{T}(z,\ell_n+\delta\gamma)
\left(\begin{array}{c}A_{n+1}\\B_{n+1}\\\end{array}\right)
=
\mathbf{T}(ik,\ell_n+\delta\gamma)
\left(\begin{array}{c}C_{n+2}\\D_{n+2}\\\end{array}\right)
\end{displaymath}
for $n=0,1,\dots,N-1$.
Equivalently, using \eqref{qua030} and \eqref{qua050}, we have 
\begin{displaymath}
\left(\begin{array}{c}C_{n+1}\\D_{n+1}\\\end{array}\right)
=
\mathbf{\Delta}(-ik\ell_n)
\mathbf{T}(ik,0)^{-1}
\mathbf{T}(z,0)
\mathbf{\Delta}(z\ell_n)
\left(\begin{array}{c}A_{n+1}\\B_{n+1}\\\end{array}\right)
\end{displaymath}
and
\begin{displaymath}
\left(\begin{array}{c}A_{n+1}\\B_{n+1}\\\end{array}\right)
=
\mathbf{\Delta}(-z\delta\gamma)
\mathbf{\Delta}(-z\ell_n)
\mathbf{T}(z,0)^{-1}
\mathbf{T}(ik,0)
\mathbf{\Delta}(ik\ell_n)
\mathbf{\Delta}(ik\delta\gamma)
\left(\begin{array}{c}C_{n+2}\\D_{n+2}\\\end{array}\right)
\;\;.
\end{displaymath}
Furthermore,  
$\mathbf{\Delta}(-z\delta\gamma)$ commutes with
$\mathbf{\Delta}(-z\ell_n)$, hence one can write:
\begin{equation}
\label{qua060}
\left(\begin{array}{c}C_{n+1}\\D_{n+1}\end{array}\right)
=
\mathbf{\Delta}(-ik(\ell_n-\delta))
\mathbf{M}\mathbf{\Delta}(ik(\ell_{n+1}-\delta))
\left(\begin{array}{c}C_{n+2}\\D_{n+2}\end{array}\right)
\end{equation}
for $n=0,\dots,N-1$, where
\begin{equation}
\label{qua070}
\mathbf{M}
=
\mathbf{\Delta}(-ik\delta)
\mathbf{T}(ik,0)^{-1}
\mathbf{T}(z,0)
\mathbf{\Delta}(-z\delta\gamma)
\mathbf{T}(z,0)^{-1}
\mathbf{T}(ik,0)
\;\;.
\end{equation}
Iterating \eqref{qua060} 
it is possible to relate $C_1$ and $D_1$ to $C_{n+1}$ and $D_{n+1}$, 
as follows:
\begin{equation}
\label{qua080}
\left( \begin{array}{c}
C \\
D_{1}
\end{array} \right)
=
\mathbf{\Delta}(ik\delta) 
\mathbf{M}^n
\mathbf{\Delta}(ik(\ell_n-\delta)) 
\left( \begin{array}{c}
C_{n+1} \\
D_{n+1}
\end{array}\right)  
\end{equation}
for $n=2,\dots,N-1$.
Then, using \eqref{qua070}, the entries of the $2\times 2$  matrix $\mathbf{M}$ are
found to be:
\begin{eqnarray}
\label{qua090}
M_{11}&=& \cos(k\delta) \cosh(z\gamma\delta) + \frac{z^2 - k^2}{2 k z} \sin(k\delta) \sinh(z\gamma\delta)   \nonumber\\
&& +i\Big(-\sin(k\delta) \cosh(z\gamma\delta) + \frac{z^2-k^2}{2 k z} \cos(k\delta) \sinh(z\gamma\delta)\Big) \nonumber\\
M_{12}&=& \frac{V}{2 k z} \sin(k\delta) \sinh(z\gamma\delta) 
          +i \frac{V}{2 k z} \cos(k\delta) \sinh(z\gamma\delta)\nonumber\\ 
M_{22}&=& M_{11}^* \nonumber\\ 
M_{21}&=& M_{12}^*, 
\end{eqnarray}
where $*$ denotes complex conjugation, and the number $z$ 
is either real or purely imaginary.
It is simple to verify that $\det(\mathbf{M})=1$.

\begin{figure}[h]
\centering
\includegraphics[width = 0.5\textwidth]{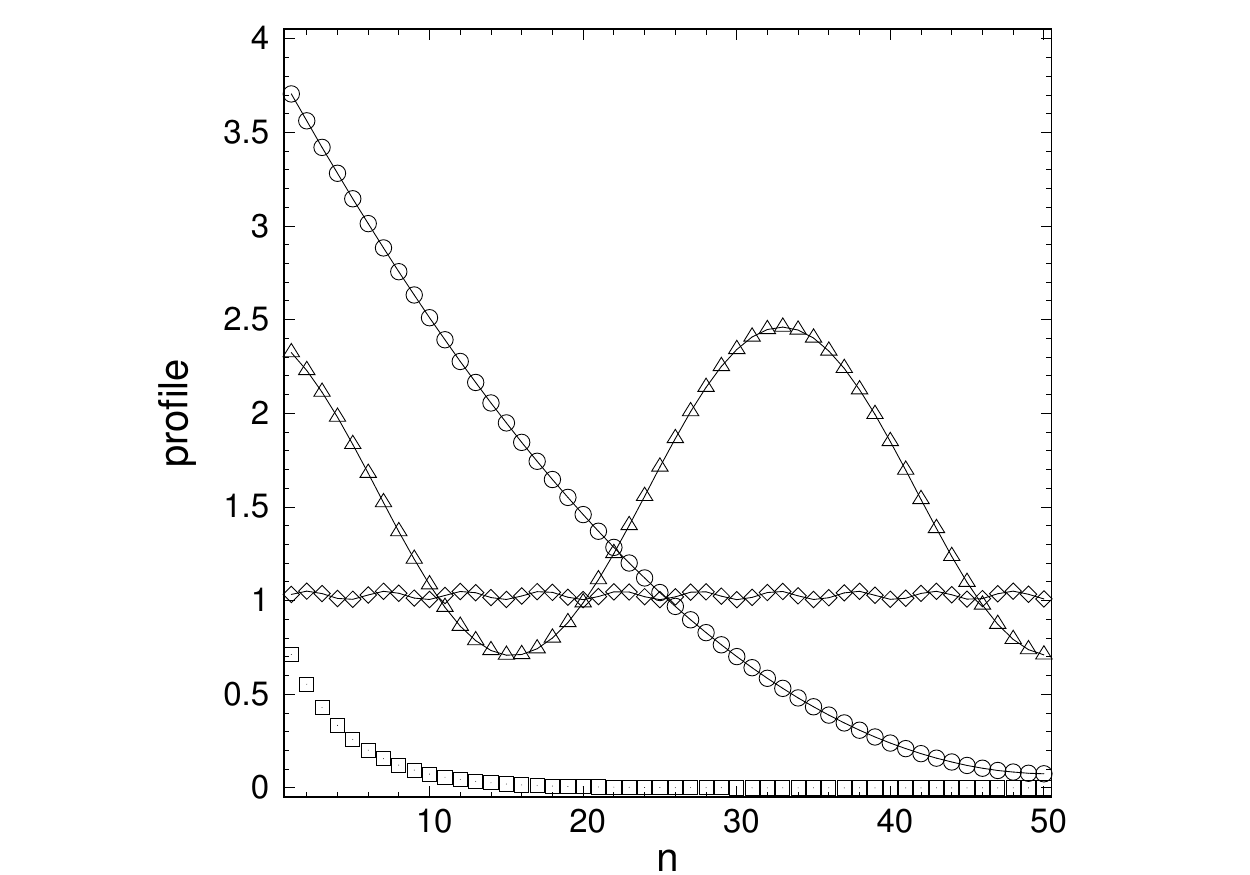} 
\caption{Profile of the mean density $\phi_n$ 
for $C=1$, $D=0$, $N=50$, $\gamma=1$, $L=10$, and $V=1$.
The different plots refer to the values of energy 
$E=0.1<E_0$ (squares), 
$E=0.5=E_0$ (circles), 
$E=0.7>E_0$ (triangles), 
and $E=10.0\gg E_0$ (diamonds).}
\label{fig:fig1}
\end{figure}

\begin{figure}[h]
\centering
\includegraphics[width = 0.45\textwidth]{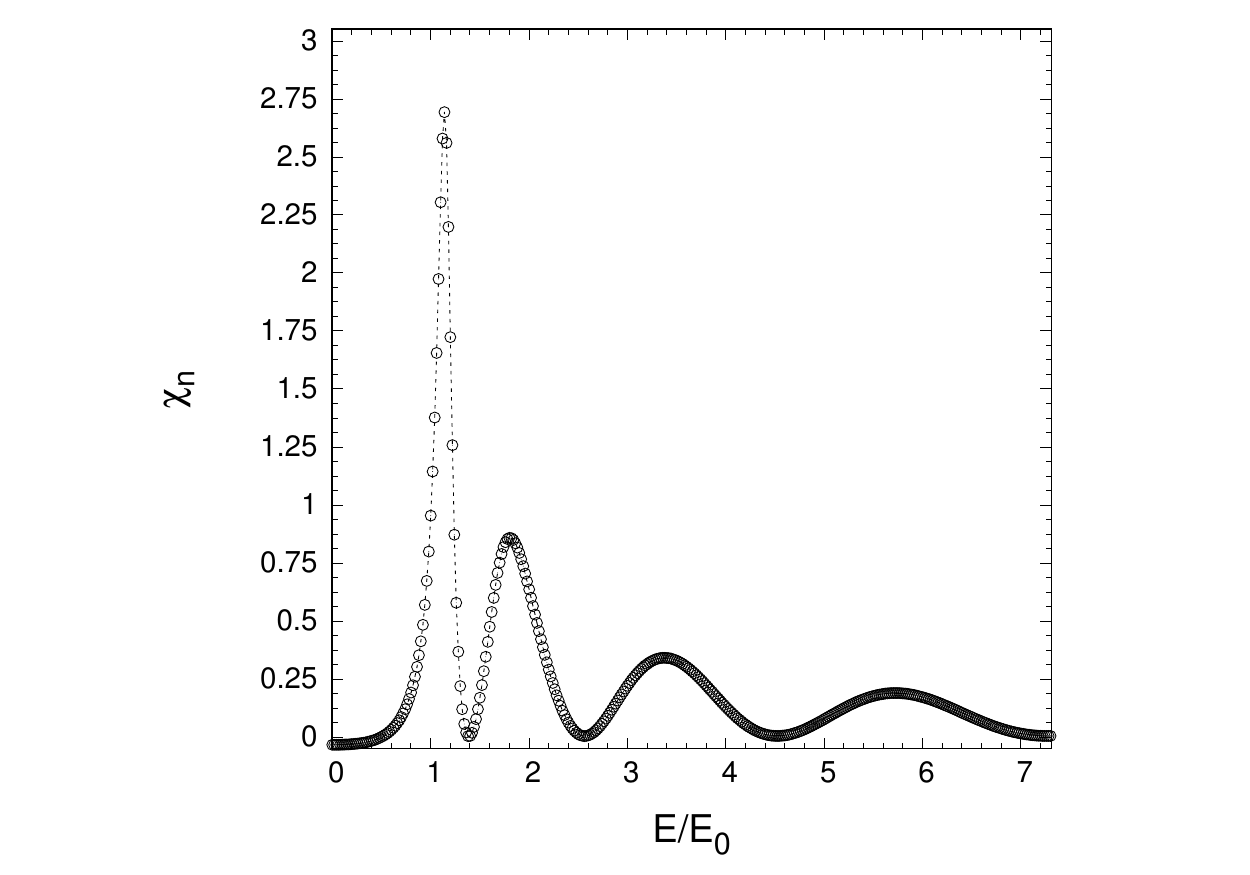} 
\includegraphics[width = 0.45\textwidth]{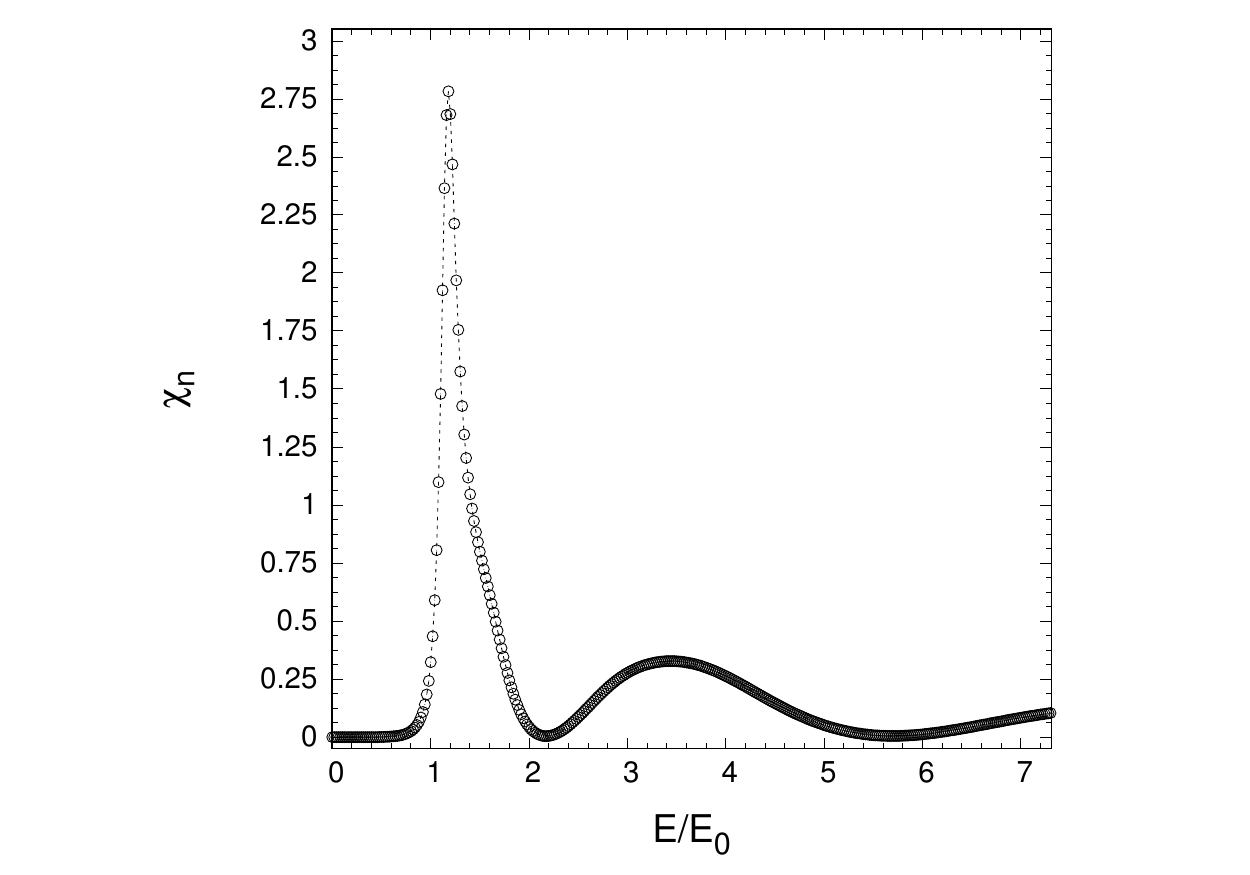} 
\caption{Interference term $\chi_n$ in \eqref{qua120} 
vs.\ the energy $E$ 
for $C=1$, $D=0$, $N=50$, $\gamma=1$, $L=10$, and $V=1$, for
$n=15$ on the left and $n=30$ on the right.}
\label{fig:fig2}
\end{figure}

We now compute the square modulus of the wave function at the points $\ell_n-\delta/2$:
\begin{equation}
\label{qua100}
|\psi_n|^2:=|\psi(\ell_n-\delta/2)|^2
=
\left|C_{n+1}\right|^2
+\left|D_{n+1}\right|^2 
+C_{n+1} D_{n+1}^* e^{2 i k (\ell_n-\delta/2)}
+C_{n+1}^* D_{n+1} e^{-2 i k (\ell_n-\delta/2)}
\end{equation}
for $n=0,\dots,N$. For $n\ge1$, this quantity is the square modulus of the wave function at the 
center of the conducting regions $[\ell_{n-1}+\gamma\delta,\ell_n]$. 
Then, each potential barrier can be associated with 
the \textit{mean density} of the two neighboring conducting regions:
\begin{equation}
\label{qua110}
\phi_n
=
\frac{1}{2}\left(\left|\psi_{n-1}\right|^2+\left|\psi_n\right|^2\right)
=
\frac{\left|C_{n}\right|^2+\left|D_{n}\right|^2
+\left|C_{n+1}\right|^2+\left|D_{n+1}\right|^2}{2}+\chi_n ~, \quad n=1,\dots,N
\end{equation}
where 
\begin{equation}
\label{qua120}
\begin{array}{rcl}
\chi_n
=S_n+S_{n+1}
\end{array}
\end{equation}
with
\begin{equation}
\label{qua125}
S_{n}
=
\frac{1}{2}
[
C_{n}D_{n}^*e^{2ik(\ell_{n-1}-\delta/2)}
+
C_{n}^*D_{n}e^{-2ik(\ell_{n-1}-\delta/2)}
]
\;\;.
\end{equation}

The quantity $\chi_n$ is due to the interference within a conducting 
region of waves 
coming from two consecutive barriers.
The mean density $\phi_n$ and the interference term $\chi_n$ profiles can be 
expressed as illustrated in the Appendix~\ref{s:appendix}. Since the 
resulting expressions are rather complex and 
analytically implicit, we plot 
the $\phi_n$ profile in 
Fig.~\ref{fig:fig1} as a function of $n$
and
the interference term $\chi_n$ in 
Fig.~\ref{fig:fig2} as a function of the energy $E$. 
Unless otherwise stated, we consider
models with $C$ a real positive constant and with $D=0$, i.e.,
cases with particles input only at the left boundary 
of the system of interest. 
As one may have expected, very large values of $E$ compared to $E_0$, 
imply that
the profile is only little affected by the material inhomogeneity. 
In particular, it is 
approximately uniform in space, 
with $\phi_n$ approximatively constant, indicating that
the transport of particles is essentially ballistic, 
since particles move almost 
exclusively in the left--to--right direction.
Nevertheless, the variety of behaviours 
shown by Figs.\ \ref{fig:fig1} and \ref{fig:fig2} indicates that the dependence of our
results on the various parameters of the quantum model is rather complex.

In our units the quantum current is defined by 
$-i/2(\psi^*\psi'-\psi^{*'}\psi)$ \cite[equation~(3.84)]{Pereira}. Between the consecutive barriers 
$n$ and $n+1$, it takes the value $k(|C_{n+1}|^2-|D_{n+1}|^2)$ 
which in a stationarity state is uniform in $n$.
Because also the plane wave is constant along the multi--barrier system, hence
the wave vector $k$ is, the uniformity of the current is expressed by:
\begin{equation}
\label{qua130}
|C|^2-|D_1|^2
=
|C_2|^2-|D_2|^2
=
\cdots
=
|C_{N+1}|^2-|D|^2
\equiv c
\;\;,
\end{equation}
so that  $k c$ is the \emph{quantum current}.
Using \eqref{qua080} with $n=N$ one can express $C_{N+1}$ 
in terms of $C$ and the current can be written as:
\begin{equation}
\label{qua135}
k c
=
\frac{k}{|(M^N)_{11}|^2}
\big[
|C|^2-|D|^2
-\big((M^N)_{12} C^* D e^{-ik(L-2\delta)}
  +(M^N)_{12}^* C D^* e^{ik(L-2\delta)}\big)
\big]
\;\;.
\end{equation}
Finally, taking $D=0$, it is natural to define the \textit{transmission coefficient} as:
\begin{equation}
\label{qua140}
S=\frac{\left|C_{N+1}\right|^2}{|C|^2}, 
\end{equation}
which is proportional to the current.

\begin{figure}[t]
\label{fig:fig3}
\centering
\includegraphics[width = 0.85\textwidth]{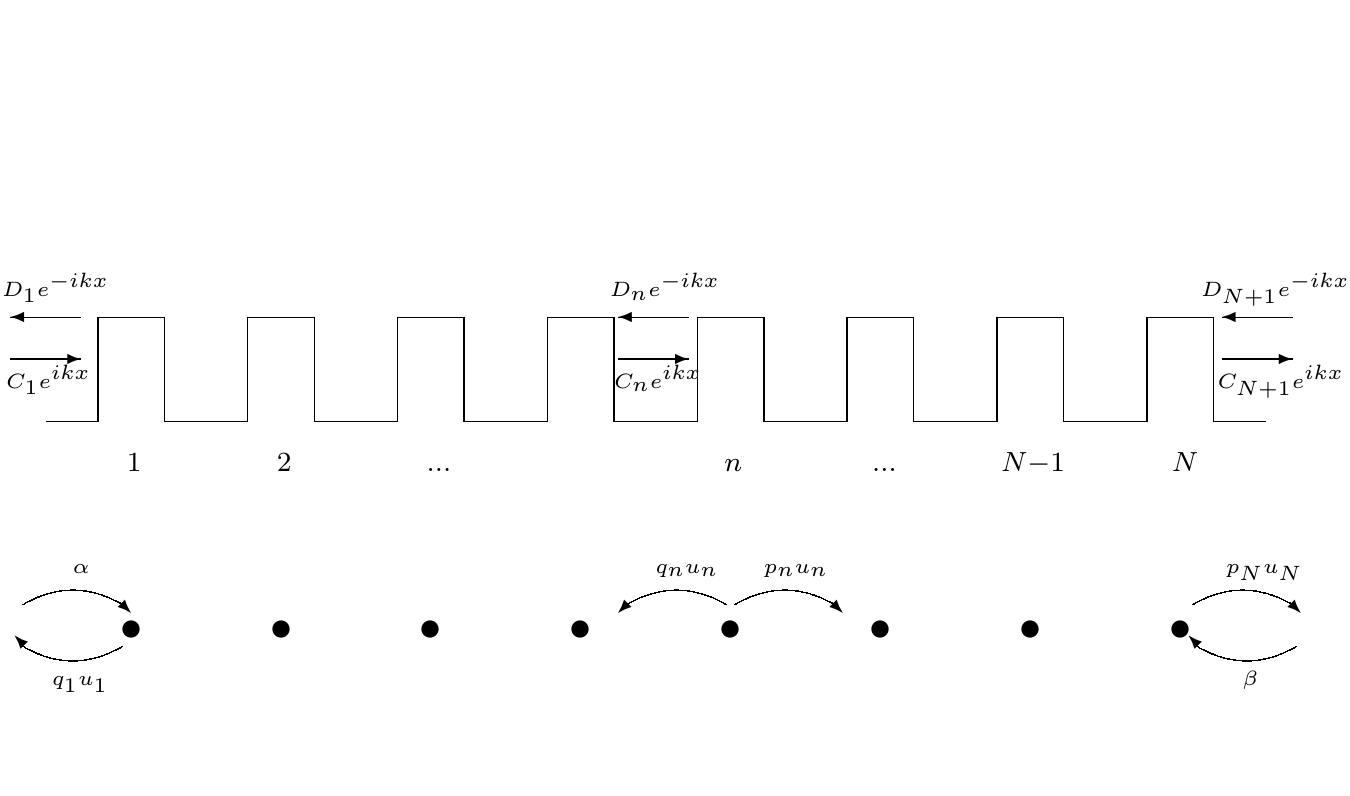} 
\caption{The equivalence between the open quantum multi--barrier system (top) and the boundary driven ZRP with independent random 
walkers (bottom) is obtained choosing the hopping probabilities $p_n, q_n$ and the injection rates $\alpha, \beta$ of the 
stochastic particle model according to Eqs. \eqref{ape190} and \eqref{ape220}.}
\end{figure}

\section{Boundary driven heterogeneous Zero Range Process}
\label{s:zrp}
\par\noindent
In this section, we show how the parameters of the quantum model illustrated in Section~\ref{s:mod} 
can be tuned so that the stationary $\phi_n$ profile coincides with the stationary density 
profile of a Zero Range Process (ZRP) 
\cite{EH2005,LS99,LMS04,CC17,CCM16,CCM16pre} 
on a 1D lattice.
As we shall see at the end of Section~\ref{s:num}, the recipe that
we give here also works for a disordered system, 
although such a case is not easily treatable in analytical terms.
The ZRP considered hereafter
consists of independent random walkers that jump to nearest neighboring sites 
with site--dependent hopping probabilities. The lattice is 
coupled, at its horizontal boundaries, to two external (infinite) particle 
reservoirs characterized by fixed injection rates, cf. Fig. \ref{fig:fig3}. 

More precisely, let us denote by $\Lambda=\{1,\dots,N\}$ a lattice with $N \ge 1$ sites,
and let the corresponding finite \emph{state} or \emph{configuration space} be denoted by 
$\Omega_N=\mathbb{N}^\Lambda$, which consists of the configurations
$\eta=(\eta_1,\dots,\eta_{N})$, where the non--negative integer $\eta_n$ represents the number of particles
at site $n$.
Given $\eta\in\Omega_N$ such that $\eta_n>0$ for some $n=1,\dots,N$, 
we let $\eta^{n,n\pm1}$ be
the configuration obtained by moving a particle from the site $n$ 
to the site $n\pm1$. We understand 
$\eta^{1,0}$ and  $\eta^{N,N+1}$ 
to be the configurations obtained by removing a particle, respectively, from
the sites $1$ and $N$.
Similarly, 
$\eta^{0,1}$ and $\eta^{N+1,N}$ denote the configurations 
obtained by adding a particle to the sites $1$ and $N$, 
respectively. The \emph{intensity} function is chosen to be 
$u(k)=k$, for $k\in\mathbb{N}$, so that the ZRP is equivalent to a superposition of
independent random walkers.

Let $p_n,q_n >0$, with $q_n=1-p_n$ denote the hopping probabilities 
on $\Lambda$, and let $\alpha,\beta>0$ be the injection rates from the left 
and the right reservoirs, respectively. The ZRP is then defined as 
the continuous time Markov jump process $\eta(t)\in\Omega_N$, $t\ge0$,
with rates
\begin{equation}
\label{ape010}
r(\eta,\eta^{0,1})=\alpha
\;\textrm{ and }\;
r(\eta,\eta^{N+1,N})=\beta
\end{equation}
for particles injection at the boundaries,
\begin{equation}
\label{ape025}
r(\eta,\eta^{n,n-1})=q_n u(\eta_n)
\;\;\textrm{ for } n=1,\dots,N
\end{equation}
for bulk left displacements, and
\begin{equation}
\label{ape028}
r(\eta,\eta^{n,n+1})=p_n u(\eta_n)
\;\;\textrm{ for } n=1,\dots,N
\end{equation}
for bulk right displacements. Then, equations \eqref{ape025} and \eqref{ape028} 
for $n=1$ and $n=N$, respectively, account 
for the particle removal at the boundaries.

The generator of this dynamics can be written as
\begin{equation}
\label{ape100}
\begin{array}{rcl}
(L_Nf)(n)
&\!\!=&\!\!
 \alpha(f(\eta^{0,1})-f(\eta))
 +
 \beta(f(\eta^{N+1,N})-f(\eta))
\\
&&\!\!
{\displaystyle
 +
 \sum_{n=1}^{N}
 [q_nu(\eta_n)(f(\eta^{n,n-1})-f(\eta))
 +p_nu(\eta_n)(f(\eta^{n,n+1})-f(\eta))]
}
\\
\end{array}
\end{equation}
for any real function $f$ on $\Omega_N$.
This means that particles hop on the lattice to the neighboring sites to the left and to the right
with rates, respectively, 
$q_n u(\eta_n)$ and $p_n u(\eta_n)$. 
The system is ``open'' in the sense that a particle hopping from the site 
$1$ or $N$ can leave the system via, respectively, a left 
or a right jump, with rates $q_1 u(\eta_n)$ and $p_N u(\eta_{N})$.
Finally, particles are injected in the system at the left and right 
boundaries with rates, respectively, $\alpha$ and $\beta$.  

The stationary measure for this process \cite{EH2005} is the product measure on the space $\Omega_N$
\begin{equation}
\label{ape020}
\mu_N(\eta)=\prod_{n=1}^{N}\nu_n(\eta_n)
\;\;\textrm{ with }\;\;
\nu_n(\eta_n)=e^{-z_n}\frac{z_n^{\eta_n}}{\eta_n!}
\end{equation}
where the real numbers 
$z_1,\dots,z_{N}$,  
called \emph{fugacities}, satisfy the equations
\cite{EH2005,LMS04}  
\begin{equation}
\label{ape050}
\begin{array}{rcl}
z_1
&\!\!=&\!\!
\alpha+q_2 z_2
\\
z_n
&\!\!=&\!\!
p_{n-1}z_{n-1}+q_{n+1}z_{n+1}
\;\;\textrm{ for }
n=2,\dots,N-1
\\
z_{N}
&\!\!=&\!\!
p_{N-1}z_{N-1}+\beta
\;\;,
\\
\end{array}
\end{equation}
which, recalling that $p_n+q_n=1$ for any $n$, can be rewritten as 
\begin{equation}
\label{ape060}
\alpha-q_1z_1
=
p_1z_1-q_2z_2
=\cdots=
p_nz_n-q_{n+1}z_{n+1}
=\cdots=
p_{N-1}z_{N-1}-q_Nz_N
=
p_Nz_n-\beta
\;\;.
\end{equation}

The main quantities of interest, in our study, are the stationary 
\emph{occupation number} profiles
\begin{equation}
\label{ape070}
\rho_n
=
\mathbb{E}_{\mu_N}[\eta_n]
=
e^{-z_n}\sum_{k=0}^\infty k\,\frac{z_n^k}{k!}=z_n
\;\;,
\end{equation}
where $\mathbb{E}_{\mu_N}$ denotes the mean computed with respect to 
the measure $\mu_N$,
and 
the stationary \emph{current}
\begin{equation}
\label{ape080}
J_{n}
=
\mathbb{E}_{\mu_N}[u(\eta_n)p_n-u(n_{n+1})q_{n+1}]
=
\mathbb{E}_{\mu_N}[\eta_np_n-\eta_{n+1}q_{n+1}]
=
p_nz_n-q_{n+1}z_{n+1}
\end{equation}
for $n=1,\dots,N$.
The stationary current 
represents the difference between the
average number of particles crossing a bond between two adjacent
sites on the lattice from the left to the right and the corresponding number hopping in the opposite direction.
From \eqref{ape050} it easily follows that 
the stationary current does not depend 
on the site $n$, as required for stationary states, 
therefore we shall simply write 
$J\equiv J_{n}=p_N z_N-\beta$.

\begin{figure}[h]
\centering
\includegraphics[width = 0.45\textwidth]{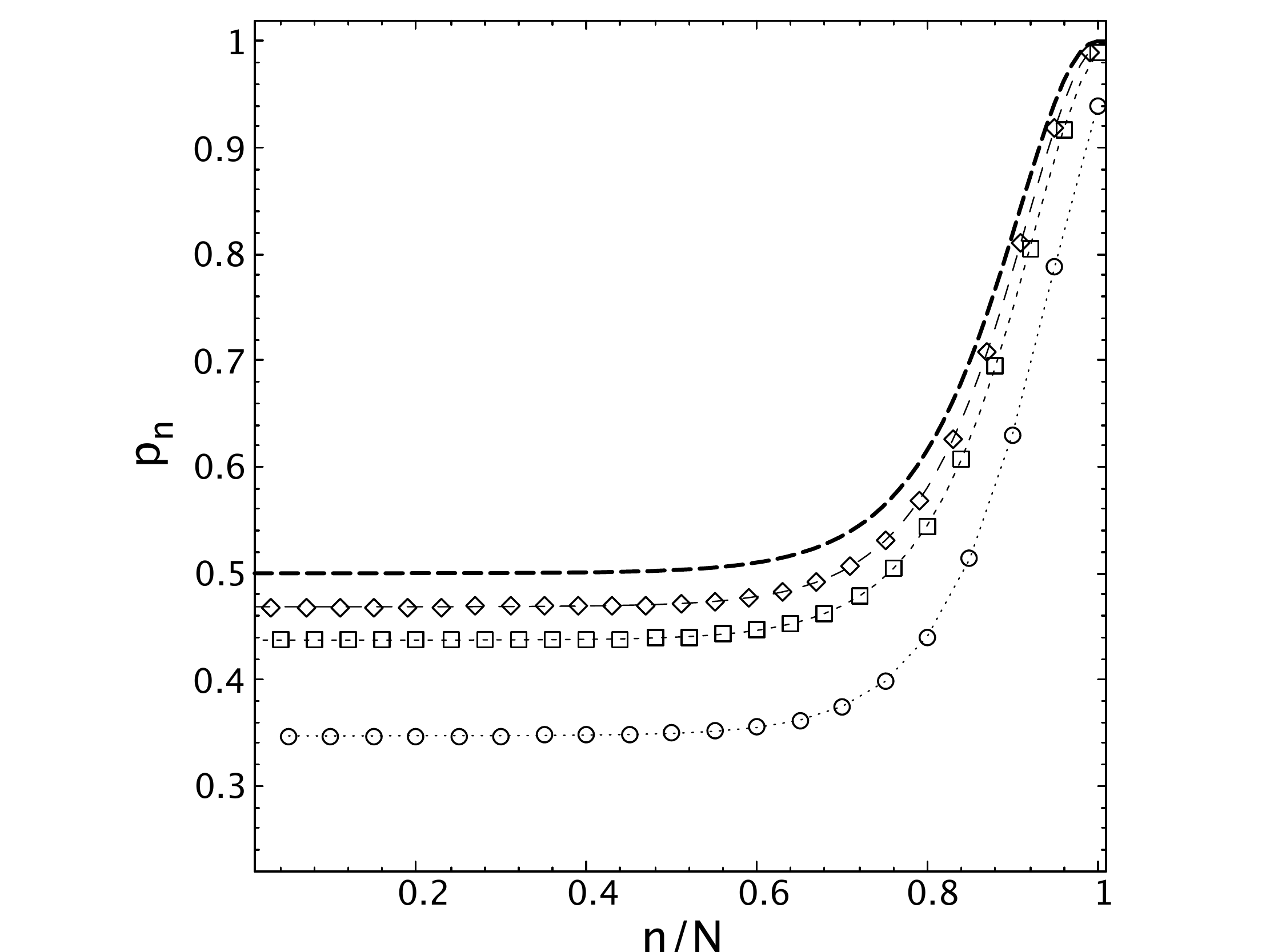} 
\includegraphics[width = 0.45\textwidth]{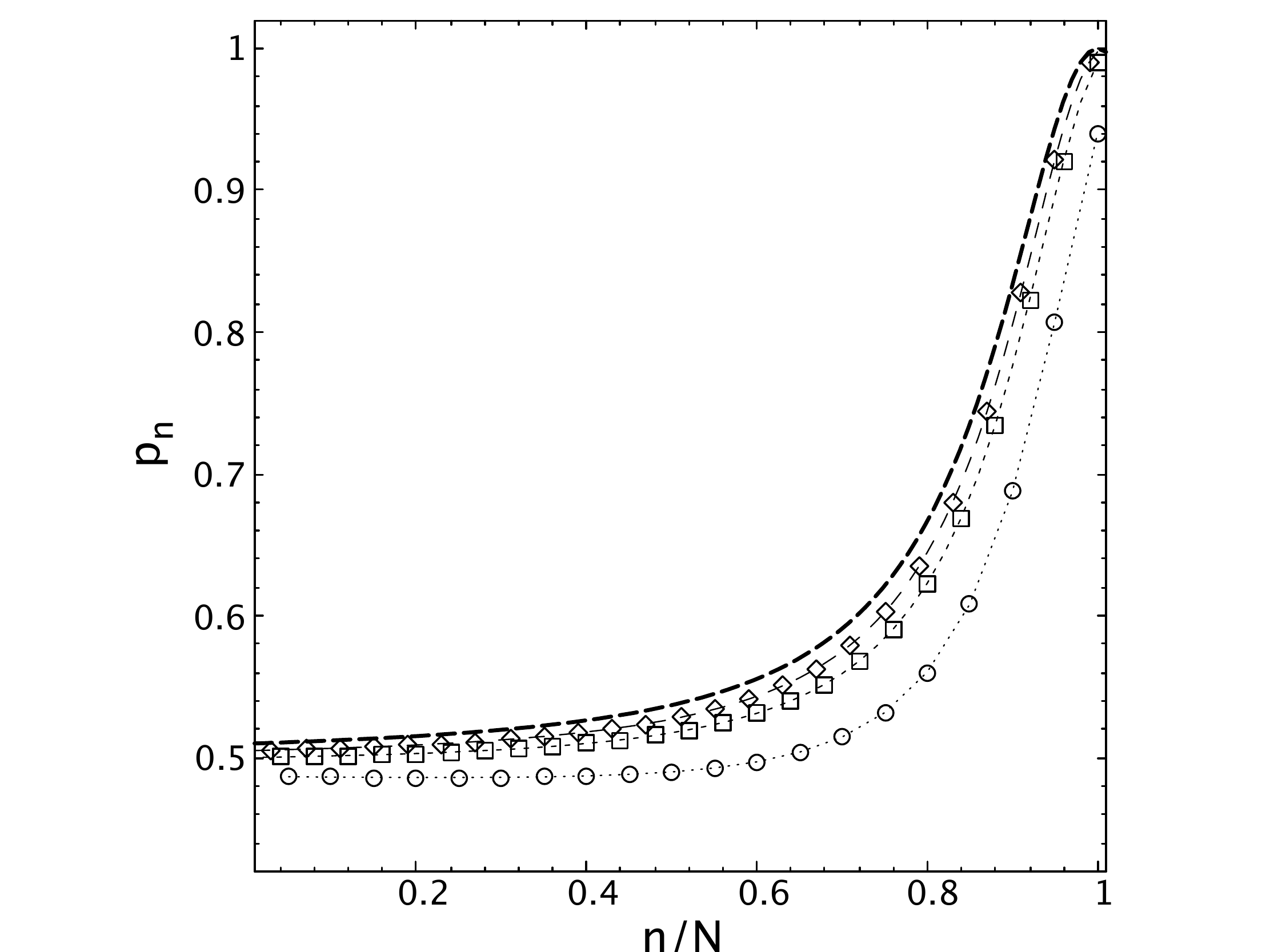} \\
\includegraphics[width = 0.45\textwidth]{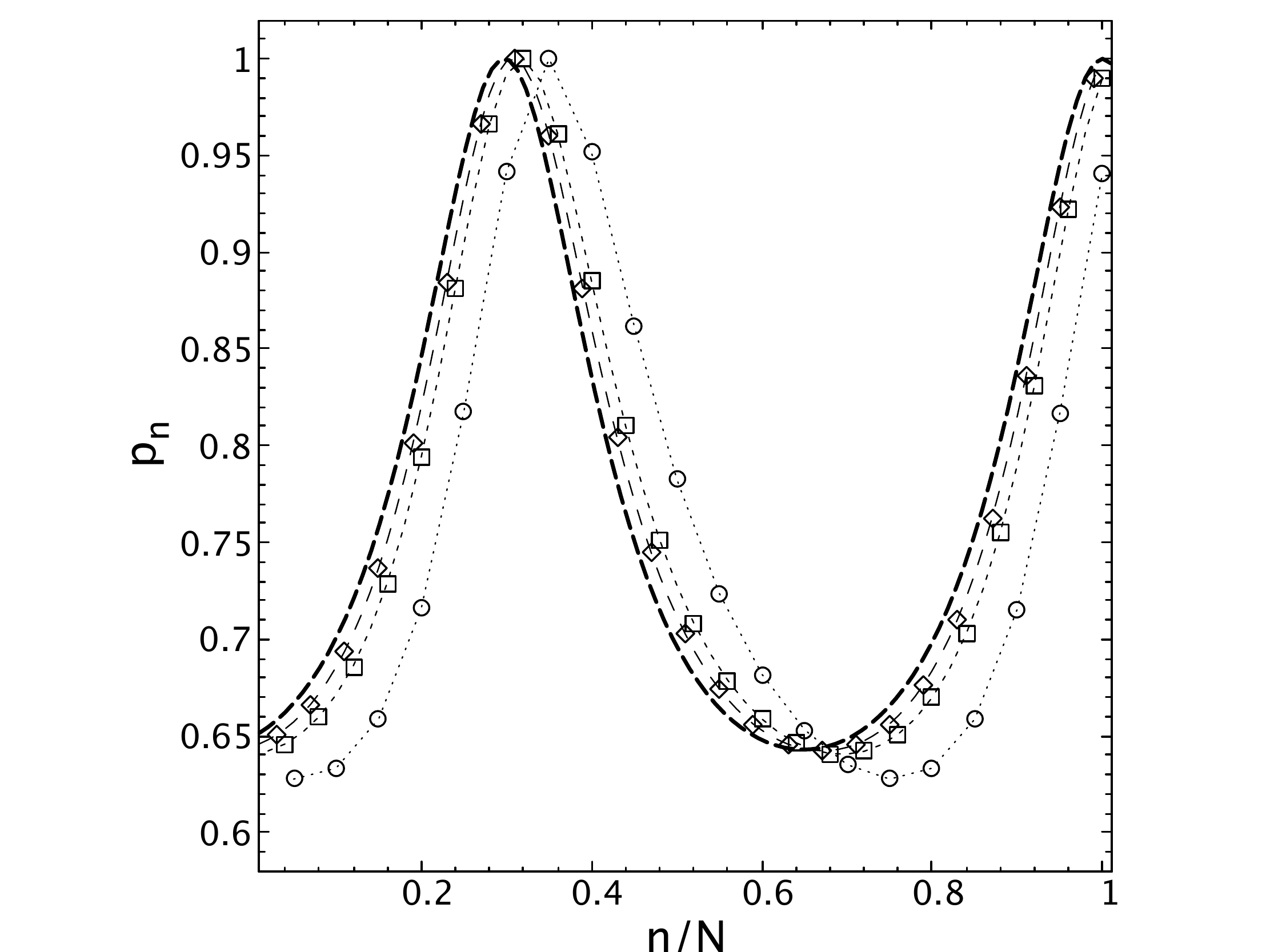} 
\includegraphics[width = 0.45\textwidth]{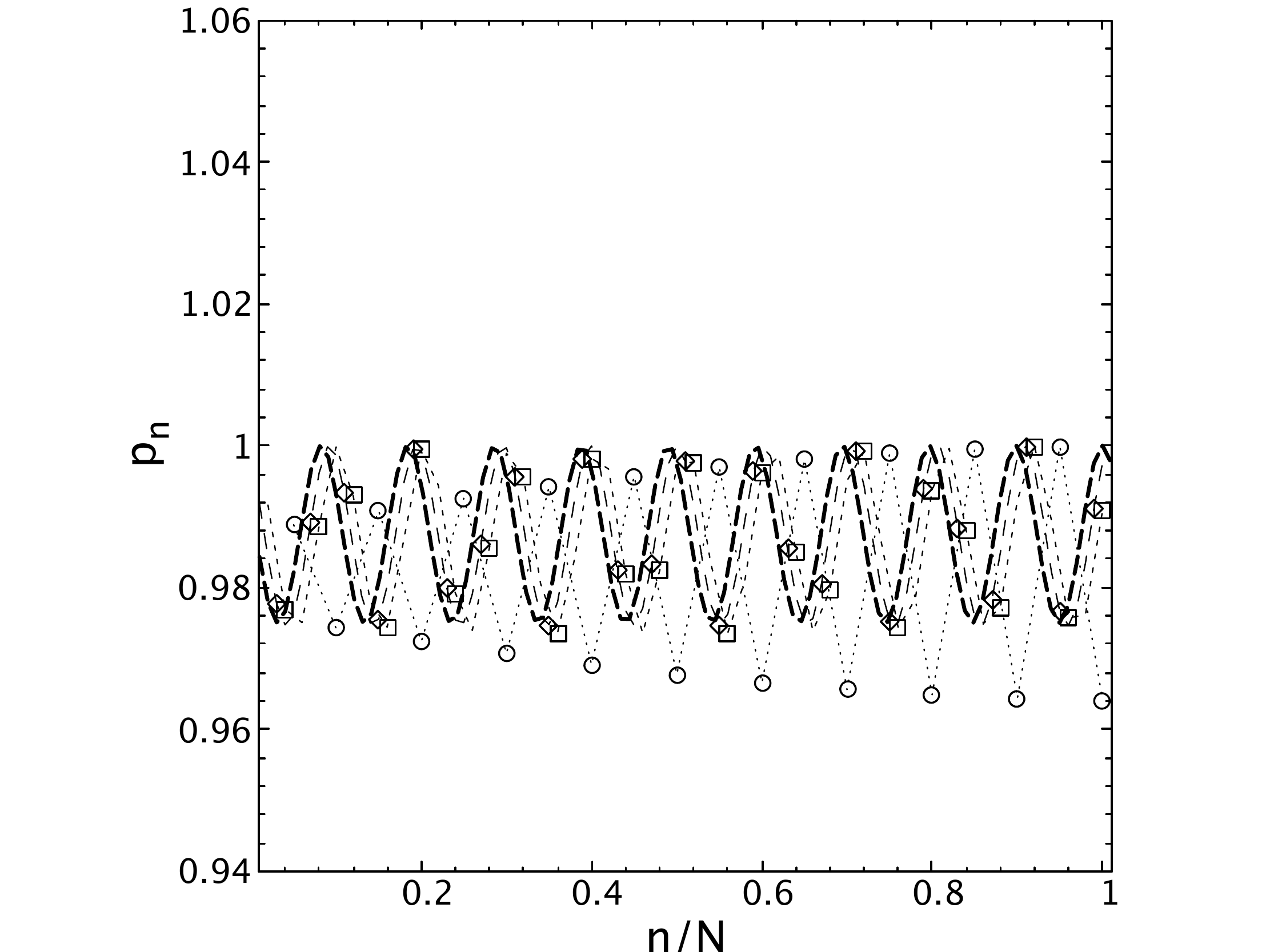} 
\caption{Hopping probabilities $p_n$ for 
for $C=5$, $D=0$, $\gamma=1$, $L=10$, $V=1$.
The energy is 
$E=0.1$ (top left panel), $E=0.5$ 
(top right panel), $E=0.7$ (bottom left panel), and $E=10.0$ (bottom right 
panel) for different values of $N$: $N=20$ (circles), $N=50$ (squares) and $N=100$ (diamonds). The thick black dashed lines, in the four panels, are 
the asymptotic hopping probabilities expressed by Eq. \eqref{pgL2}.}
\label{fig:fig4}
\end{figure}

The equivalence of the stationary profiles of the stochastic ZRP and of the 
quantum Kronig--Penney model is now obtained tuning the parameters as follows. 
First, introduce 
\begin{equation}
\label{ape190}
I_n
=
|C_{n+1}|^2
+
S_{n+1}
+
|D_{n}|^2
+
S_n ~, \quad n=1,\dots,N
\end{equation}
with $S_n$ defined by Eq. \eqref{qua125}.
Then, in order to match the density profiles of the quantum process with those of 
the ZRP,
when the parameters of the quantum process are given, we take 
\begin{equation}
\label{ape200}
p_n
=
\frac{|C_{n+1}|^2
+S_{n+1}}
 {I_n}
\;\;\textrm{ and }\;\;
q_n
=
\frac{|D_{n}|^2
+S_n}
 {I_n}
\;\;, \quad n=1,\dots,N
\end{equation}
as hopping probabilities, and
\begin{equation}
\label{ape220}
\alpha
=
|C|^2
+
S_1
\;\;\textrm{ and }\;\;
\beta
=
|D|^2
+S_{N+1}
, 
\end{equation}
as the left and right injection rates of the ZRP.
The idea underlying this identification is that $\left|\psi_n\right|^2$, namely the square modulus 
of the wave function evaluated at the center of the conducting region on 
the right of the $n$--th barrier ({\em i.e.}\ at $\ell_n-\delta/2$) yields
two contributions: one, corresponding to $\left|C_{n+1}\right|^2+S_{n+1}$, associated with the 
average rate of the right jump of a particle from the $n$--th site and another, given by $\left|D_{n+1}\right|^2+S_{n+1}$, 
associated with the average rate of the left jump from the $(n+1)$--th site. 
It is important to note that, the rules \eqref{ape200} and \eqref{ape220} 
give non--negative hopping 
probabilities that sum to 1 when $|C_n|^2+S_n>0$ and $|D_n|^2+S_n>0$ 
for all $n=0,1,\dots,N+1$. As explained
in the Appendix~\ref{s:appendix}, this is the case 
in the $\gamma,L$--continuum limit \cite{CNP15} with $D=0$, namely, for $N$ 
large enough when all the other parameters are kept fixed.

The values of the hopping probabilities $p_n$ for several different values of the energy are plotted in Fig.~\ref{fig:fig4}. Note that in the various panels the thick black dashed lines indicate the behavior of the probability in the $\gamma,L$--continuum limit considered in the Appendix~\ref{s:appendix}, see Eq. \eqref{pgL2}.

The top left panel of this figure shows that, for $E<E_0$, such probabilities are practically constant 
and smaller than $1/2$ in most of the space. 
As the energy grows, the hopping probabilities develop oscillations till they settle about the
value 1.

Equations \eqref{ape200} and \eqref{ape220} imply  
\begin{equation}
\label{ape230}
\alpha-I_1q_1=|C|^2-|D_1|^2,\;
I_Np_N-\beta=|C_{N+1}|^2-|D|^2
\end{equation}
and
\begin{equation}
\label{ape240}
p_nI_n-q_{n+1}I_{n+1}=|C_{n+1}|^2-|D_{n+1}|^2
\end{equation}
for $n=1,\dots,N-1$.
Using the conservation of the quantum current, Eq. \eqref{qua130}, in Eqs. \eqref{ape230} and \eqref{ape240}, and by comparing with Eq. \eqref{ape060}, one thus obtains $z_n=I_n$ for $n=1,\dots,N$, and the equivalence between the quantum current and the current \eqref{ape080} of the stochastic model is established.
Moreover, since Eq. \eqref{qua130} yields 
$|C_n|^2+|D_{n+1}|^2=|C_{n+1}|^2+|D_n|^2$, one realizes that equations \eqref{qua110} 
and \eqref{ape070} imply $\phi_n=\rho_n$ for every $n$, which 
is the equivalence of the quantum mean density profile 
and the stationary occupation profile of the stochastic particle system.

We conclude this section recalling that, as also noted at the end of 
Section~\ref{s:mod}, the quantum  
current \eqref{qua130} reduces to the transmission coefficient 
in \eqref{qua140}, when $D=0$ and $C=1$. Thus, the recipe given in 
Section~\ref{s:zrp}.
allows us to interpret the transport properties of the quantum model in terms of the 
stationary current of a boundary driven random walk on a 1D lattice \cite{EH2005}.

\section{Discussion of profiles in stationary states}
\label{s:num}
\par\noindent
In this section we compare the exact expression 
\eqref{qua110} of $\phi_n$ and $\rho_n$ with the profile obtained from 
Monte Carlo simulations of the ZRP model. We also compare
the current $J$ of the stochastic process, obtained from Monte Carlo simulations, 
with the $c$ of \eqref{qua130}, that is the quantum current divided by $k$. 
To do that, we take the parameters values prescribed 
by equations \eqref{ape200} and \eqref{ape220}.
We further take  
$C=5$, $D=0$, $N=50$, $\gamma=1$, $L=10$, and $V=1$, which 
yields $E_0=0.5$, cf.\ \eqref{zero}.
Because $D=0$, given an energy $E>0$, $N$ must be correspondingly large, 
cf.\ comment below \eqref{ape220} 
and the Appendix~\ref{s:appendix}.

The numerical simulations are performed as follows: the process 
starts with zero particles, but the injection of particles at the 
left boundary quickly makes the lattice populated. After a sufficiently 
large time, the particles distribution reaches a stationary state. 
At stationarity, the occupation number profile is measured 
with two different methods, that are mathematically equivalent, but numerically
independent:
i) the number of particles at each site 
is averaged collecting its values at time intervals larger than the decorrelation time,
that is of the order of the number $N$ of sites;
ii) the total number of particles jumping from a generic site $n$ 
to the left is computed and, at the 
end of the simulation, is divided by the total 
time, that is given by the sum of exponentially distributed time intervals 
between two consecutive jumps, and by the probability 
$q_n$ to perform the left jump. Using \eqref{ape080} also this ratio 
should yield $z_n=\rho_n$. 

The match between these two independent calculations demonstrates the good accuracy of our numerical simulations,
which is important, since in the sequel we shall discuss the 
disordered case, for which no analytical results are currently available.

\begin{figure}[t!]
\centering
\includegraphics[width = 0.45\textwidth]{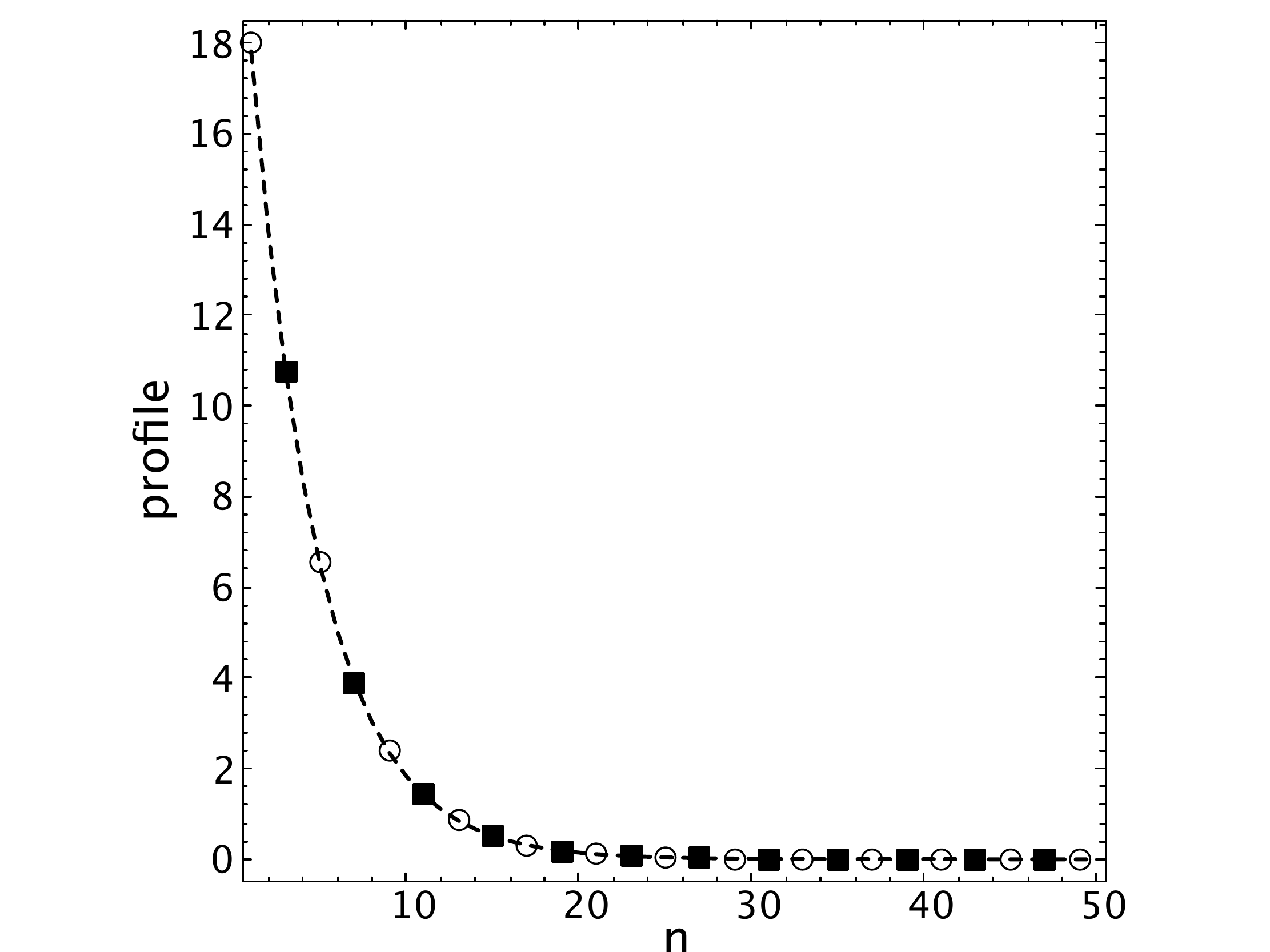} 
\includegraphics[width = 0.45\textwidth]{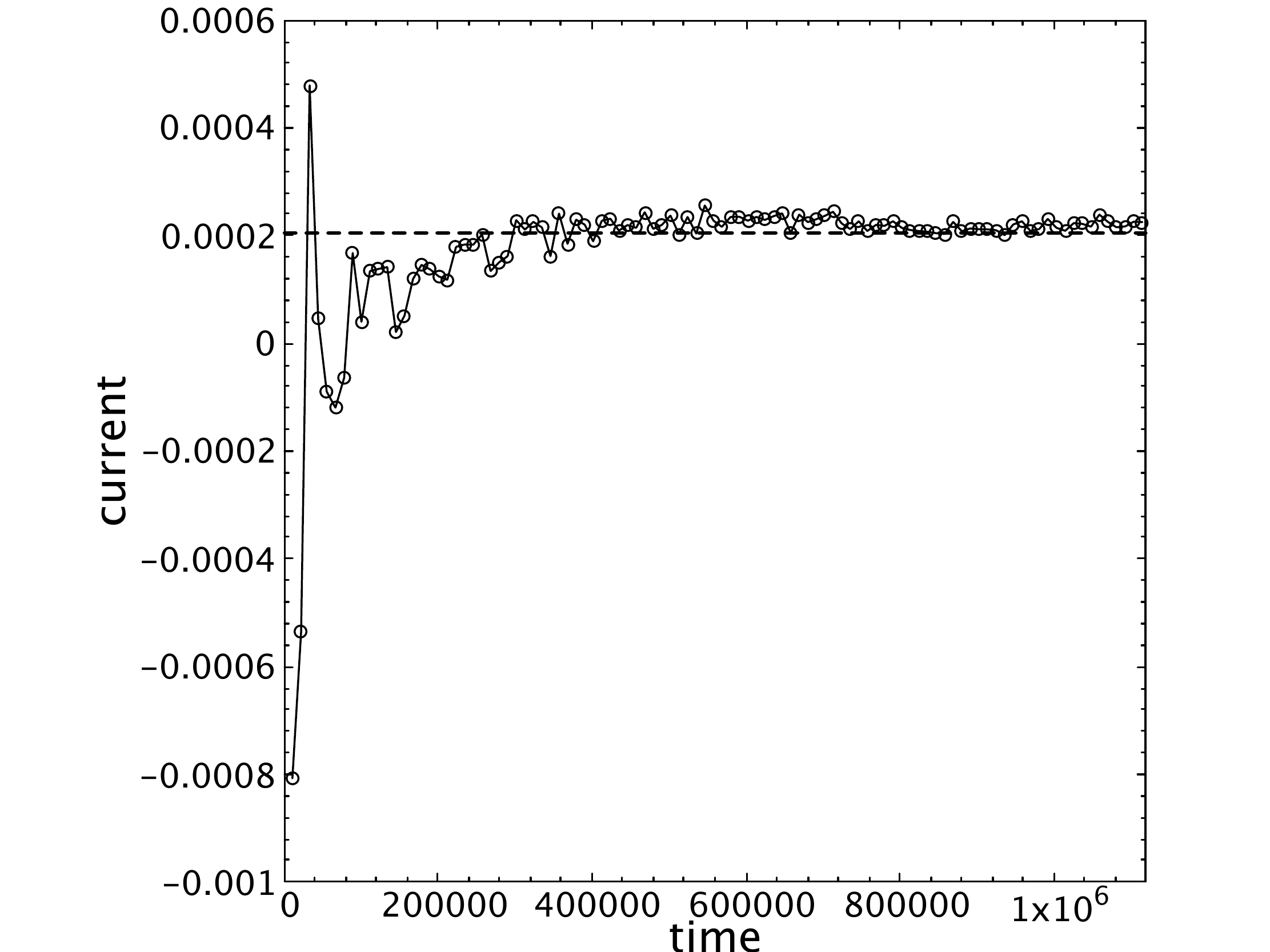} 
\caption{\textit{Left panel}: 
Profile of particles density $\phi_n$ of the quantum 
multi--barrier system given by \eqref{qua110} (dashed line),
and occupation number profile of the ZRP process,
with hopping probabilities \eqref{ape200}, 
obtained via MC simulations,  by measuring the stationary site occupation (open circles) as well as the ratio of the stationary hopping rate to the left to the corresponding hopping probability  $q_n$ (filled squares),
for 
$C=5$, $D=0$, $N=50$, $\gamma=1$, $L=10$, $V=1$, 
$E=0.1$. 
\textit{Right panel}: MC measure of the current in the ZRP model  
as a function of time $t$ (empty circles) compared to the 
theoretical value of the quantum transmission coefficient multiplied by $\left|C\right|^2$, see Eqs. \eqref{qua135} 
and \eqref{qua140} (dashed line).}
\label{fig:fig5}
\end{figure}

\begin{figure}[h!]
\centering
\includegraphics[width = 0.45\textwidth]{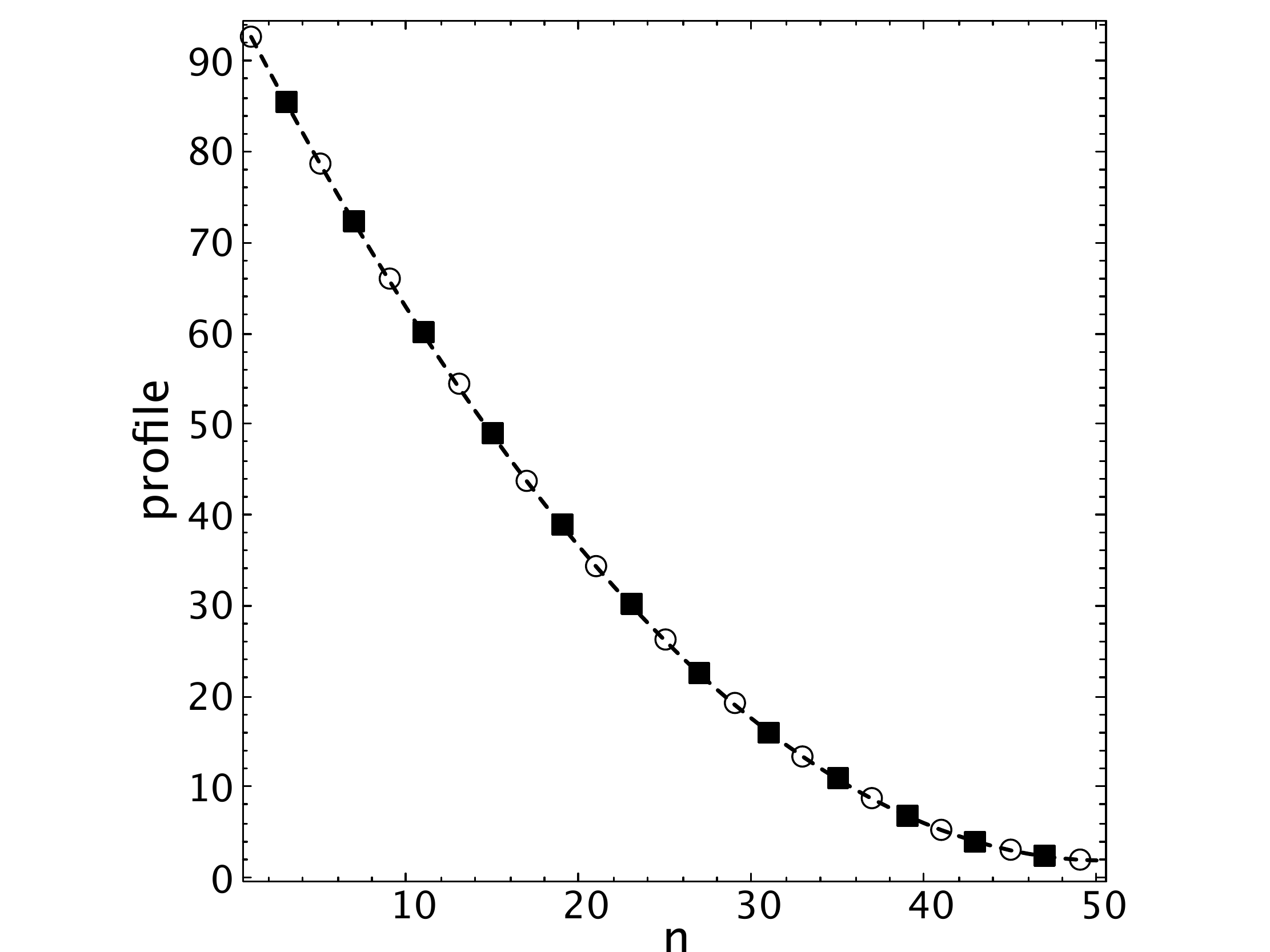} 
\includegraphics[width = 0.45\textwidth]{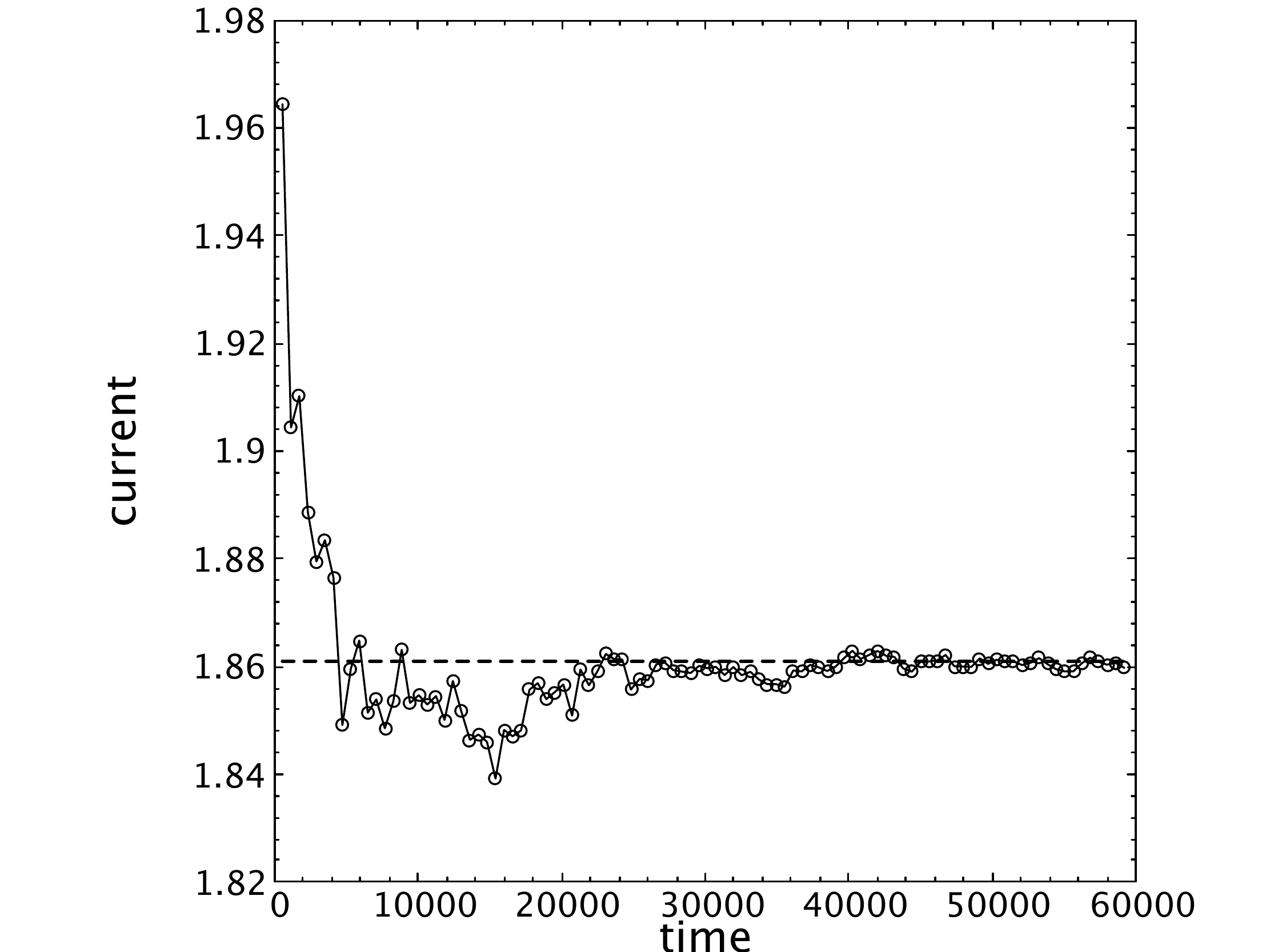} 
\caption{As in Fig.~\ref{fig:fig5} for $E=0.5$.}
\label{fig:fig6a}
\end{figure}

\begin{figure}[h!]
\centering
\includegraphics[width = 0.45\textwidth]{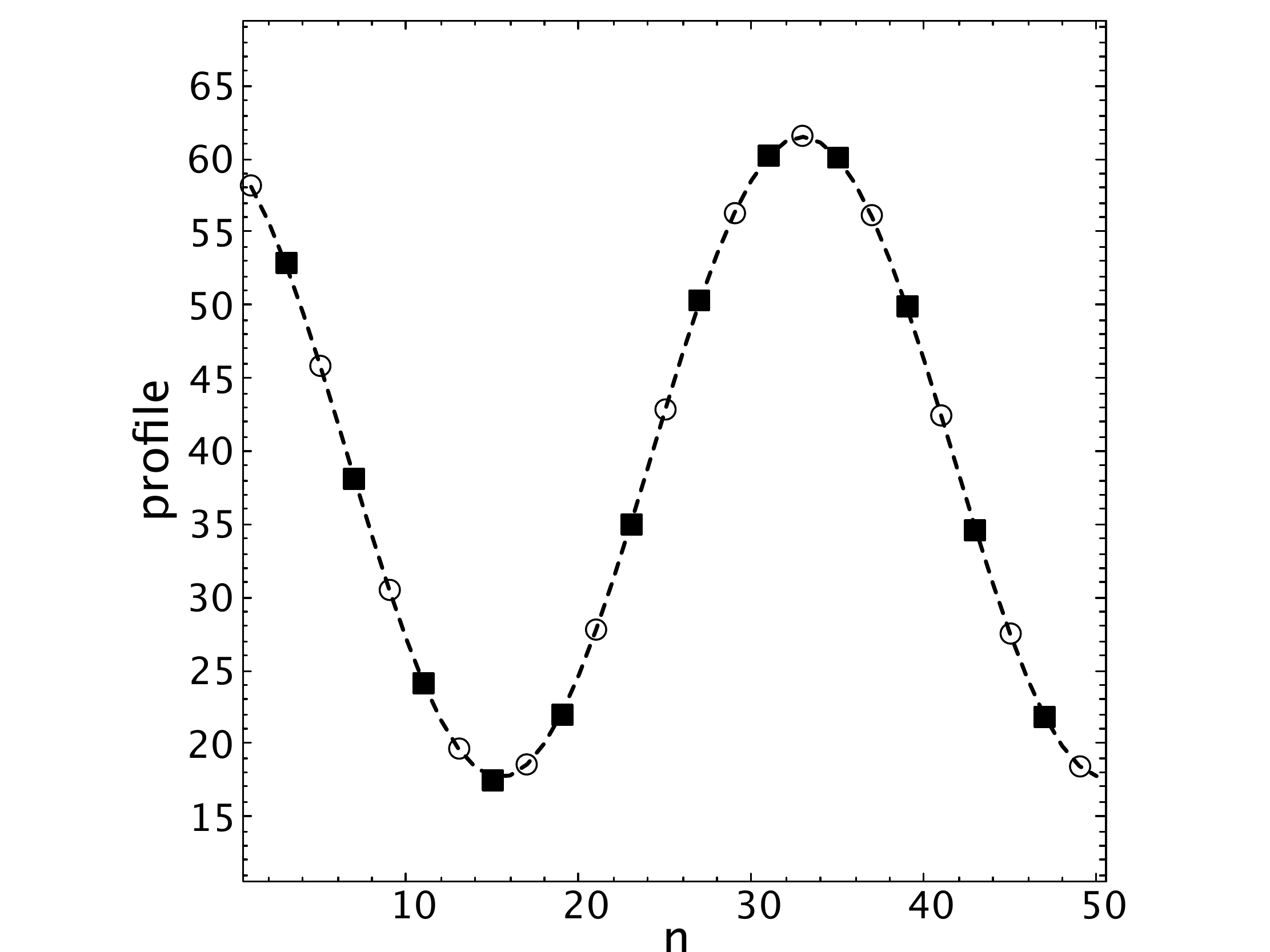} 
\includegraphics[width = 0.45\textwidth]{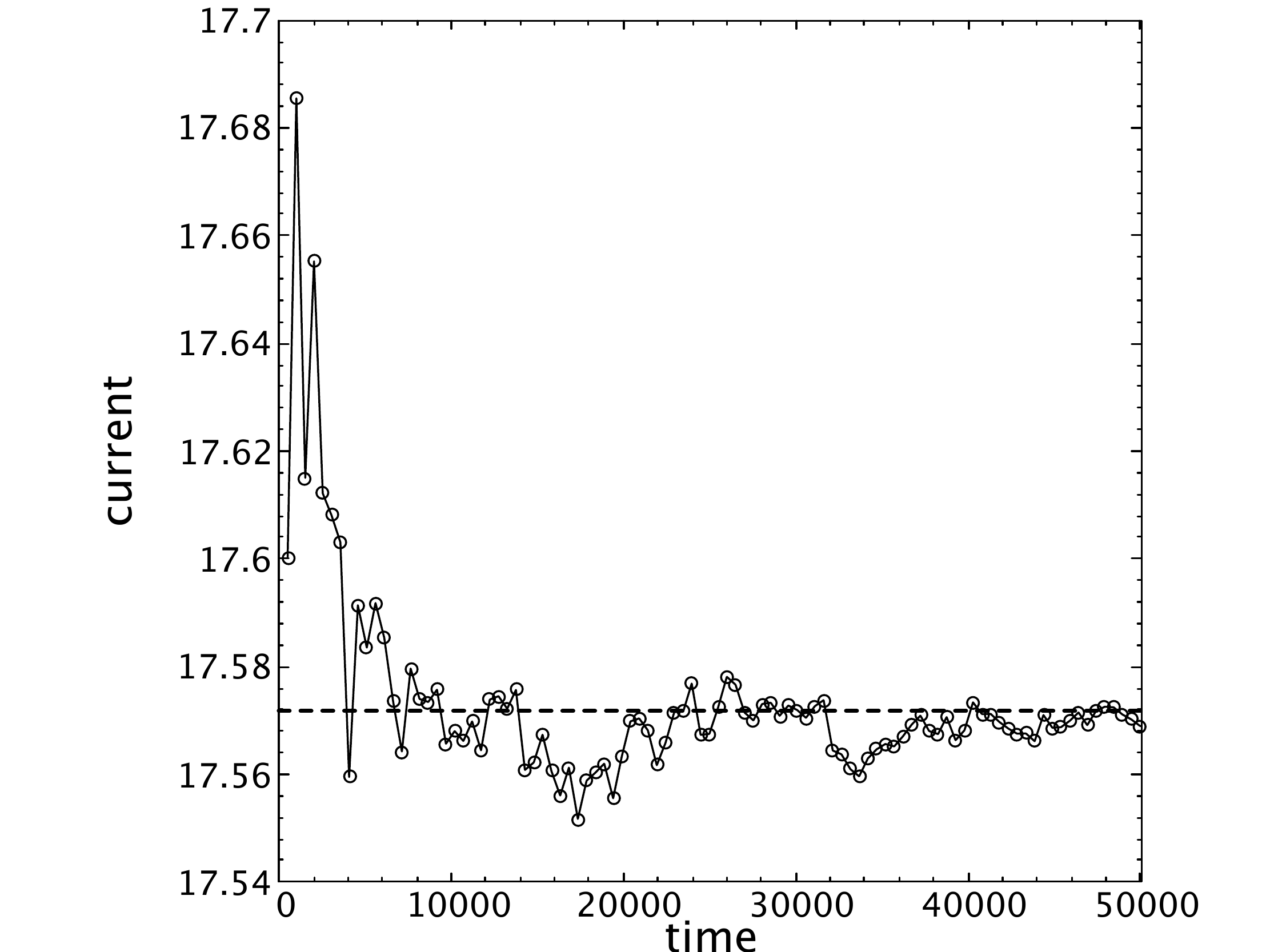} 
\caption{As in Fig.~\ref{fig:fig5} for $E=0.7$.}
\label{fig:fig6}
\end{figure}

\begin{figure}[h!]
\centering
\includegraphics[width = 0.45\textwidth]{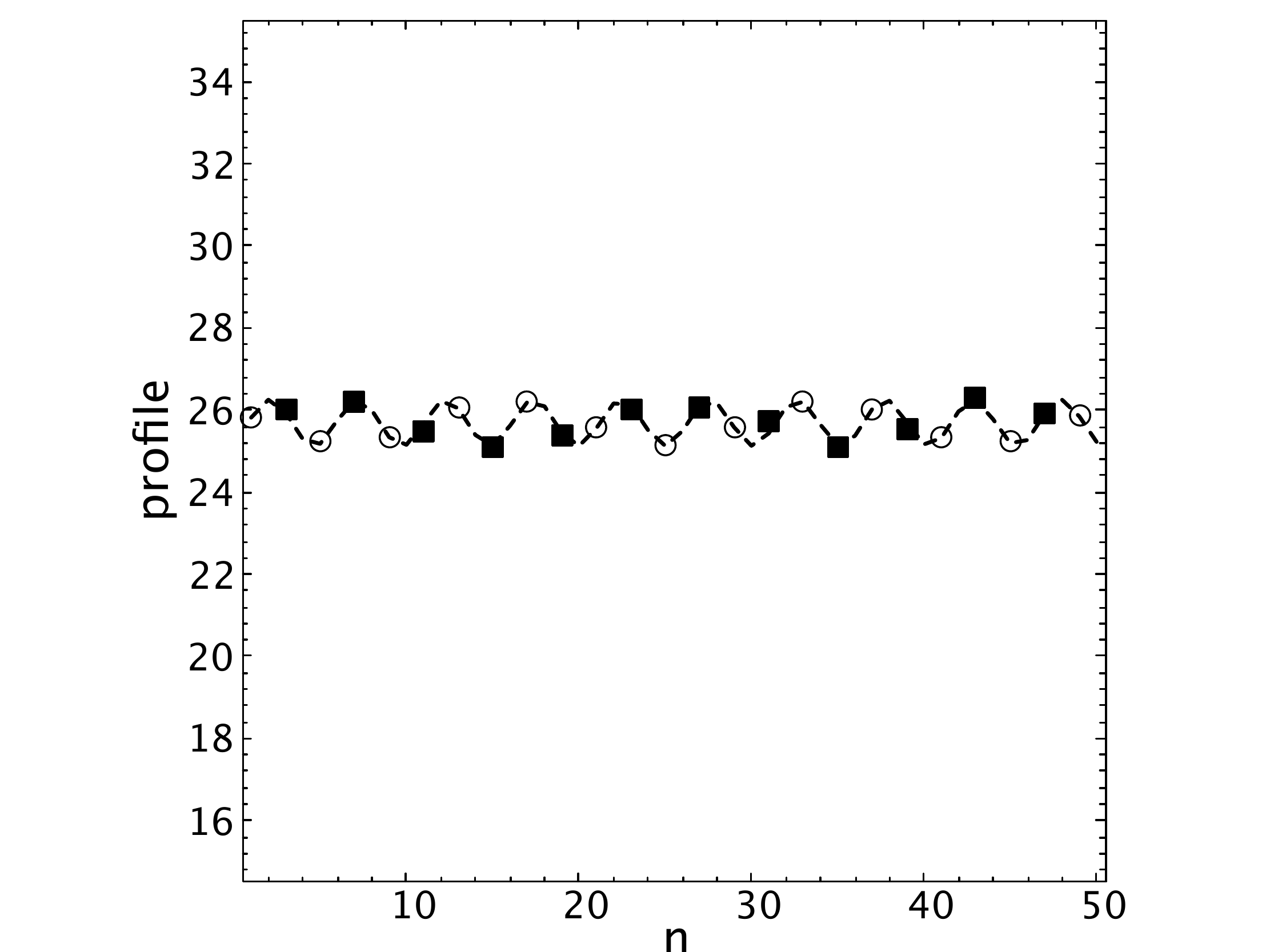} 
\includegraphics[width = 0.45\textwidth]{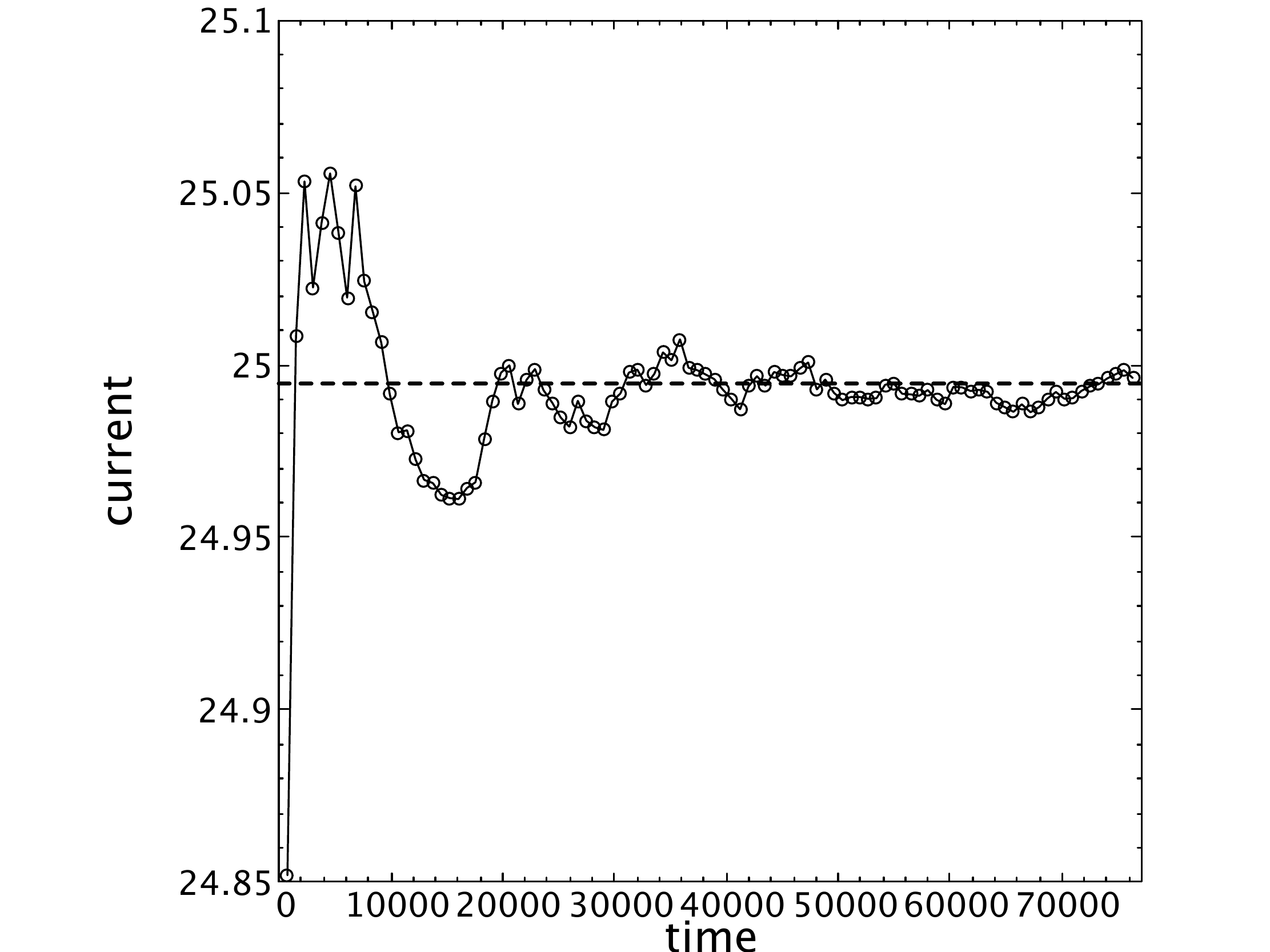} 
\caption{As in Fig.~\ref{fig:fig5} for $E=10$.}
\label{fig:fig7}
\end{figure}

We shall now
distinguish the regimes with energy $E < E_0$, and cases with 
energy $E > E_0$, as suggested in \cite[Remark 4.4]{CNP15}. In our study $E_0=1/2$. 

Case $E=0.1<E_0$, \textit{monotonic decay}:
The top left panel of Fig.~\ref{fig:fig2} shows that the 
$p_n$ is constant in most of the system positions, and smaller than $1/2$. 
At the right end of the system a sudden increase takes 
place.
The occupation number profile is, instead, monotonically decreasing, see Fig.~\ref{fig:fig5},
and convex. 

Case $E=E_0=0.5$, \textit{transition behavior}:
The $p_n$ profile in this case 
is shown in the top right panel of Fig.~\ref{fig:fig2}, and it
turns out to be monotonically increasing.
The occupation number profile, plotted in 
Fig.~\ref{fig:fig6a}, is monotonically decreasing, but the total 
mass on the lattice is larger with respect to the previous case. 
This can be ascribed to the fact that in that case the injection rate $\alpha$ from the left reservoir has increased with respect to the case with $E=0.1$.\begin{figure}[t!]
\centering
\includegraphics[width = 0.45\textwidth]{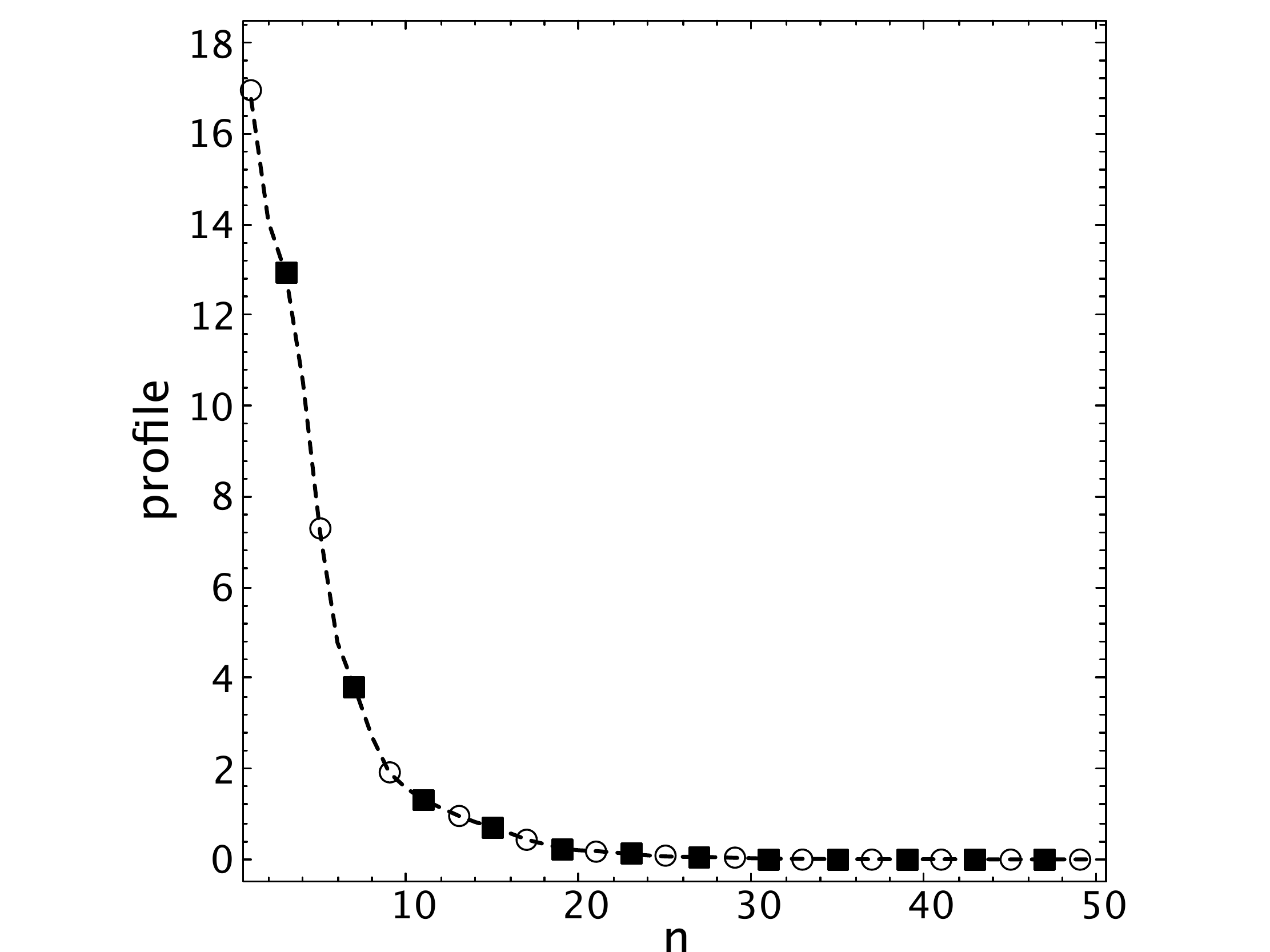} 
\includegraphics[width = 0.45\textwidth]{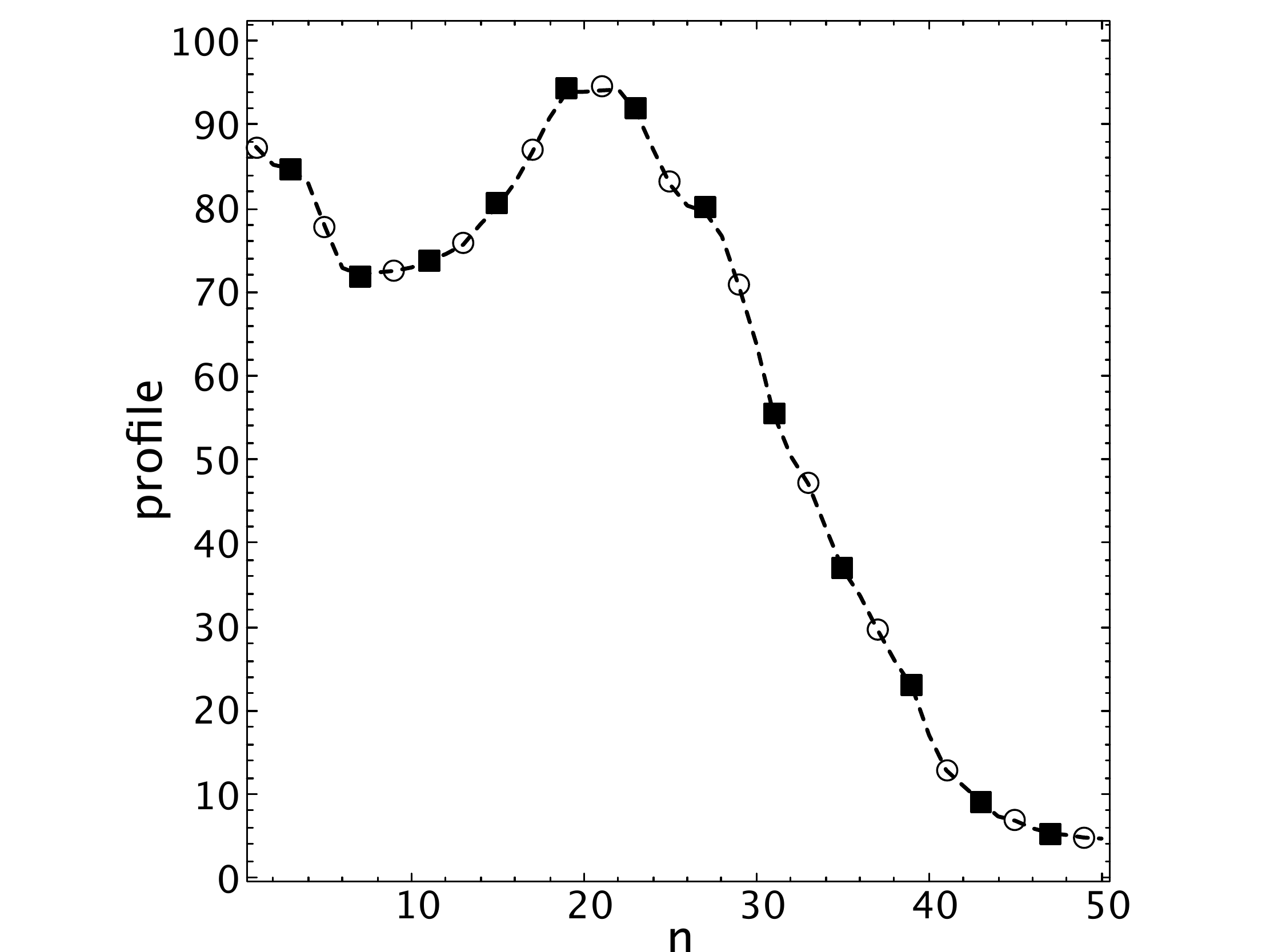} \\
\includegraphics[width = 0.45\textwidth]{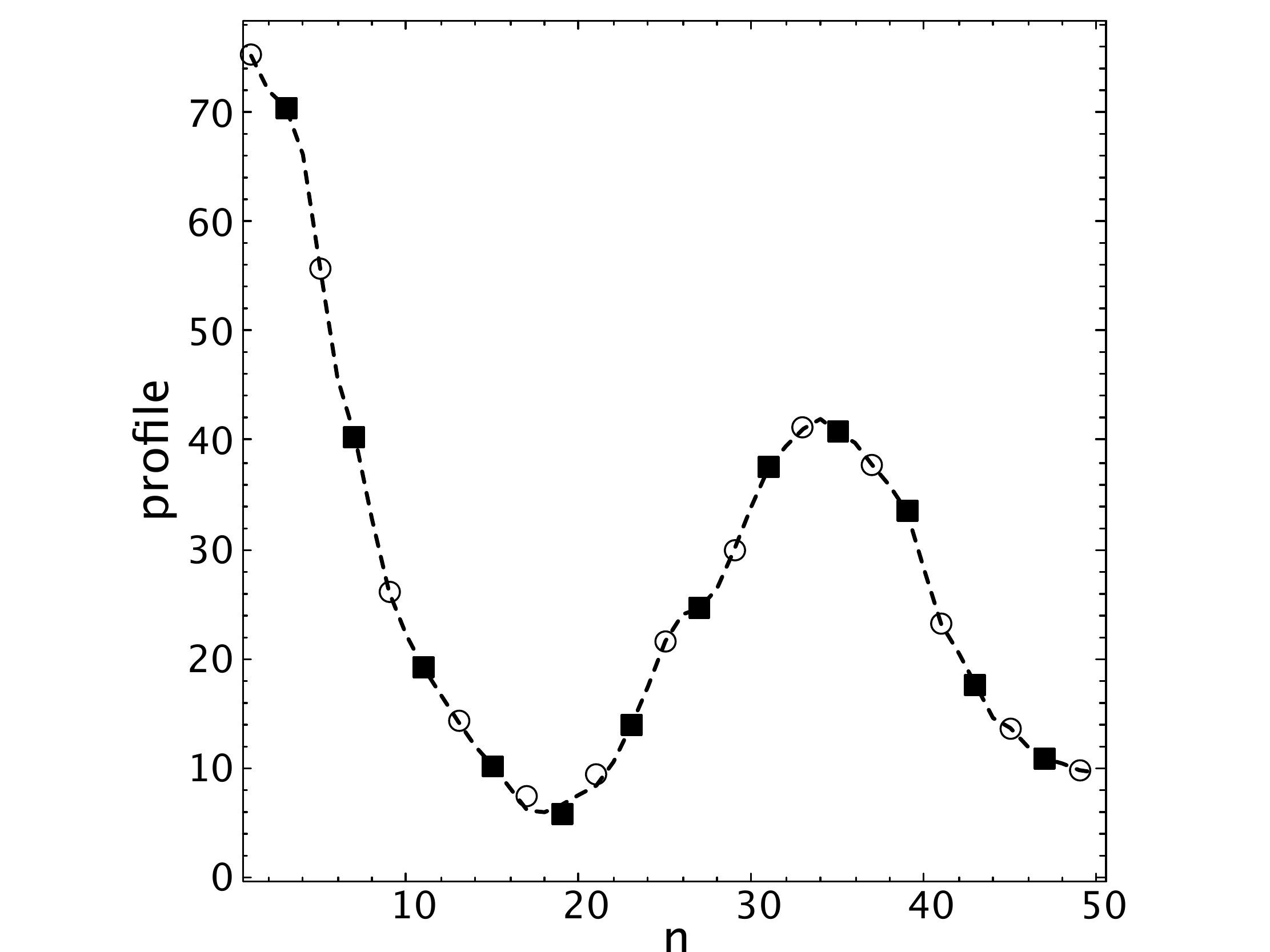} 
\includegraphics[width = 0.45\textwidth]{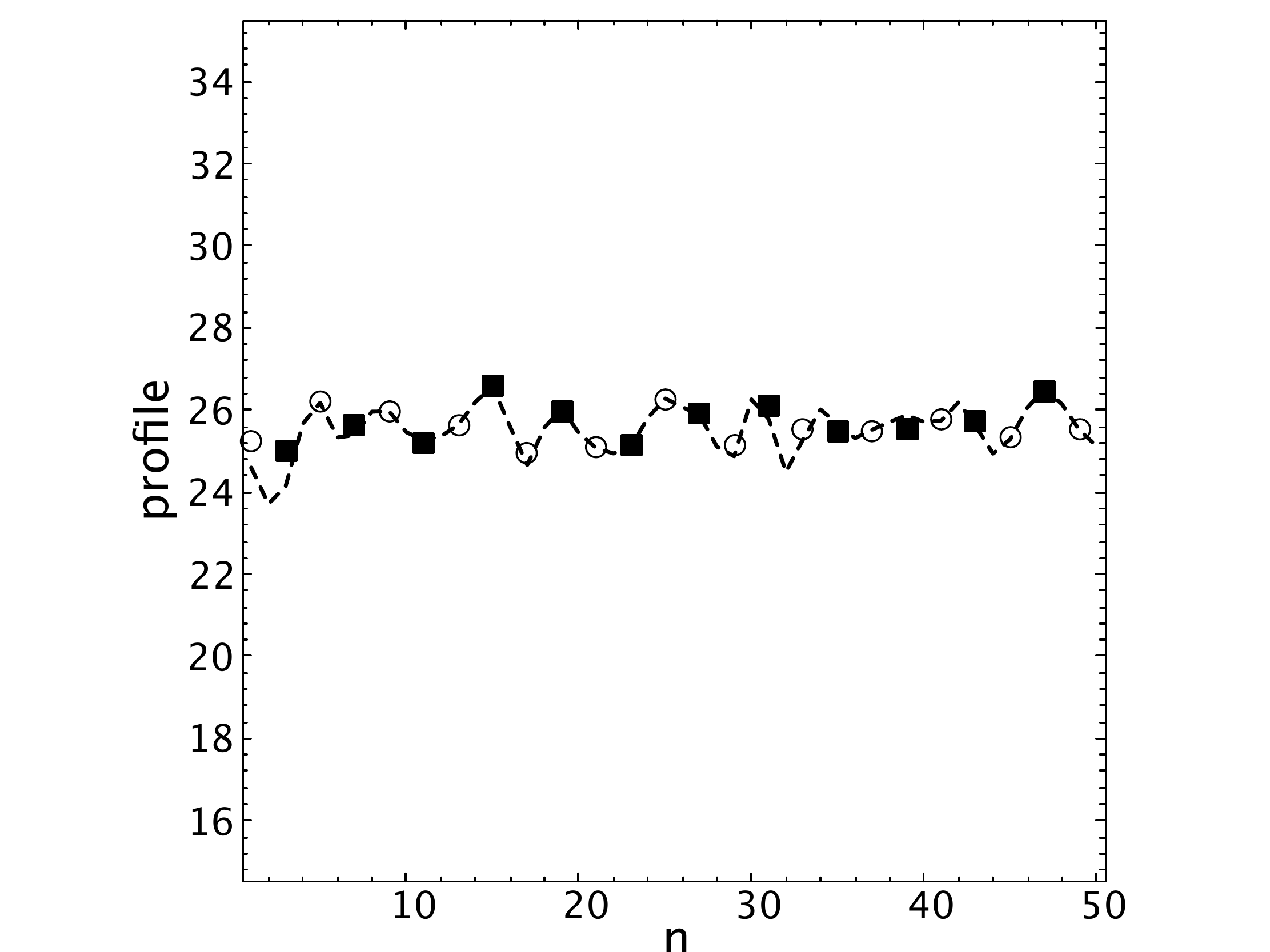} 
\caption{Profile of $\phi_n$ for a single realization
of the quantum disordered multi--barrier system given by \eqref{qua110} (dashed line),
and occupation number profile of the ZRP process 
with hopping probabilities \eqref{ape200} 
obtained via MC simulations,  by measuring the stationary site occupation (open circles) as well as the ratio of the stationary hopping rate to the left to the corresponding hopping probability  $q_n$ (filled squares),
for 
$C=5$, $D=0$, $N=50$, $\gamma=1$, $L=10$, $V=1$ and with different values of energy: $E=0.1$ (top left panel), $E=0.5$ (top right panel), $E=0.7$ (bottom left panel) and $E=10$ (bottom right panel).}
\label{fig:fig8}
\end{figure}

Case $E=0.7>E_0$, \textit{oscillatory behavior}:
The probability profile $p_n$ 
is shown in the bottom left panel of Fig.~\ref{fig:fig2} and 
it is an oscillatory function. 
As a consequence, the occupation number profile shown in 
in Fig.~\ref{fig:fig6} is also an oscillating function.

Case $E=10\gg E_0$, \textit{ballistic behavior}:
The transition probability profile $p_n$ is shown in the bottom panel 
of Fig.~\ref{fig:fig2} and it is 
very close to $1$. 
Therefore, the stochastic model is made of
particles undergoing a strongly 
asymmetric random walk on the lattice. Fig.~\ref{fig:fig7} shows the 
occupation number profile which is, indeed, almost uniform. 

Finally,
we investigate numerically the equivalence between a disordered version of 
the Kronig--Penney model introduced in 
\cite{Luna09} and a boundary driven ZRP with random hopping probabilities, 
cf.\ e.g.\ \cite{Koukkous99,Ferrari07}.
The quantum disordered model is constructed by drawing, first, the random 
variables $\lambda_n, \delta_n$ from a uniform 
distribution, and then by rescaling them, in order to fulfill the 
two constraints
\begin{equation}
\label{constr1}
\sum_{n=1}^N (\lambda_n - \gamma \delta_n) =  0
\;\;\textrm{ and }\;\;
\sum_{n=1}^N (\lambda_n +  \delta_n) = L
\end{equation}
with $\gamma,L$ fixed. 
Remarkably, the correspondence method developed in Section~\ref{s:zrp}  still holds here,
as long as the parameters $p_n$ and $q_n$ defined by Eq.~\eqref{ape200} are positive and smaller than one.
This result is established via numerical simulations; in fact we can not guarantee
it in the random case, since the corresponding product of random matrices does not lead
to an expression like  \eqref{app020}.

The results of the MC simulations are reported 
for a single realization of the disordered quantum system
in Fig.~\ref{fig:fig8}, which shows a striking equivalence between the
quantum and the stochastic processes here investigated. The black dashed line 
displayed in the various panels of Fig. \ref{fig:fig8} corresponds to the exact quantum
profile $\phi_n$ in Eq.~\eqref{qua110}, obtained by using the transfer matrix formalism 
discussed in Section~\ref{s:mod}, whereas the filled and empty symbols denote the results 
of the MC simulations based on the two alternative methods outlined before.

\begin{figure}[t!]
\centering
\includegraphics[width = 0.32\textwidth]{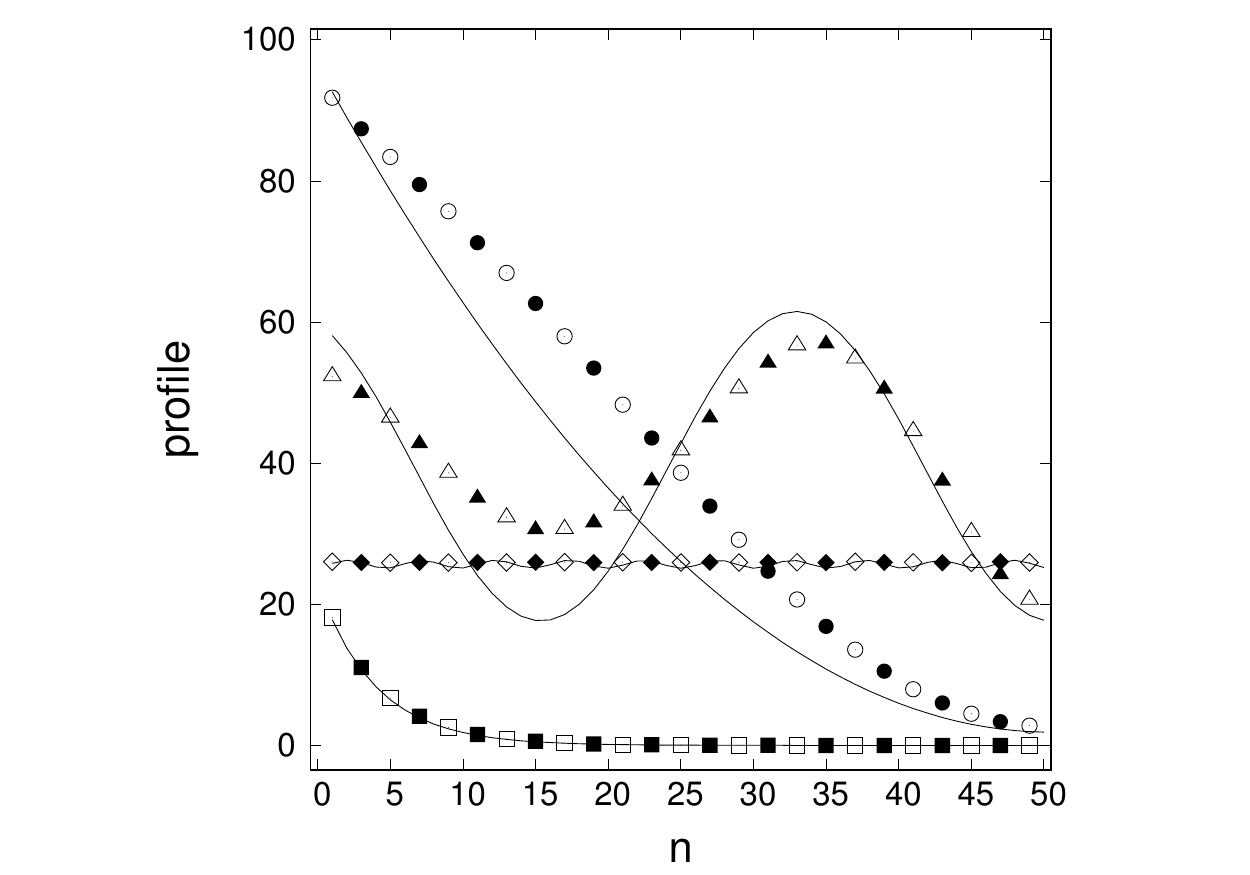} 
\includegraphics[width = 0.32\textwidth]{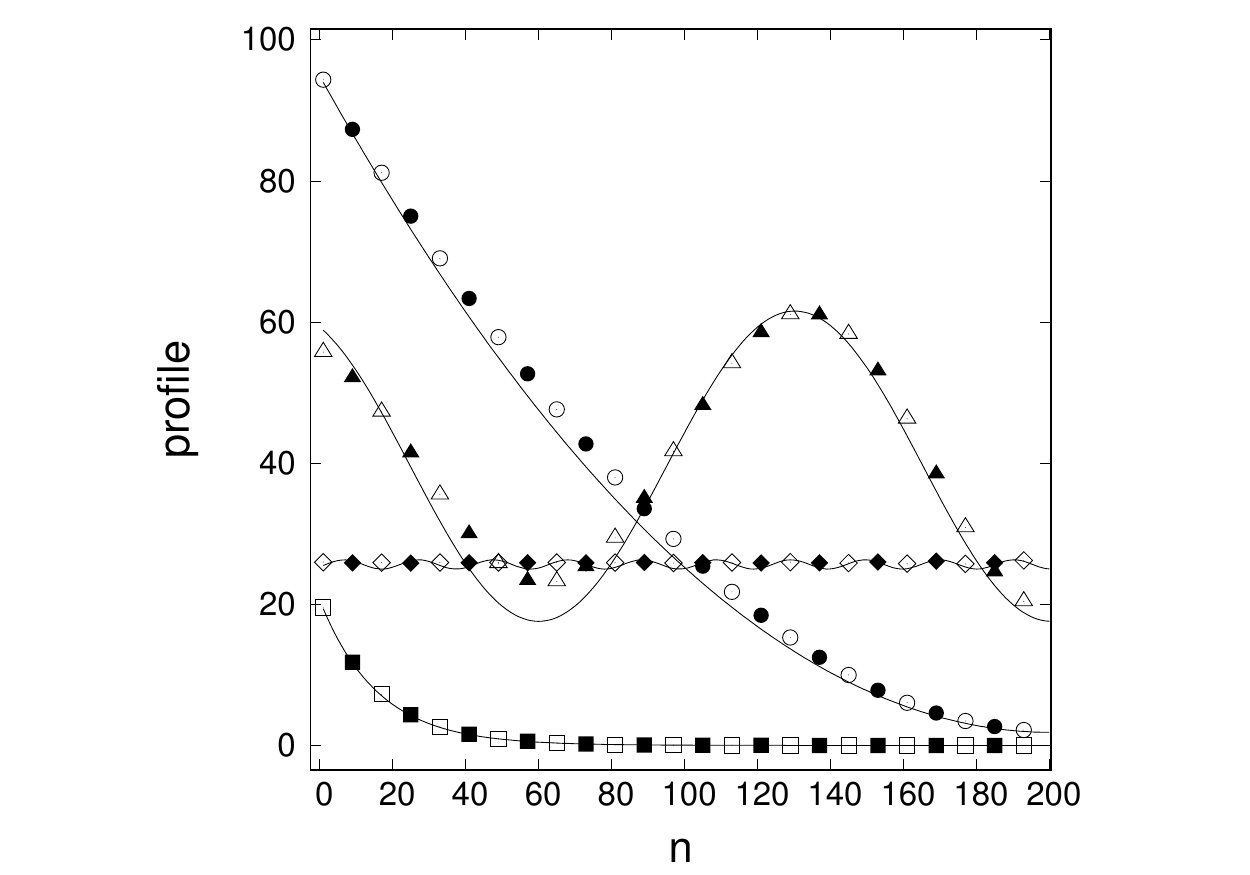} 
\includegraphics[width = 0.32\textwidth]{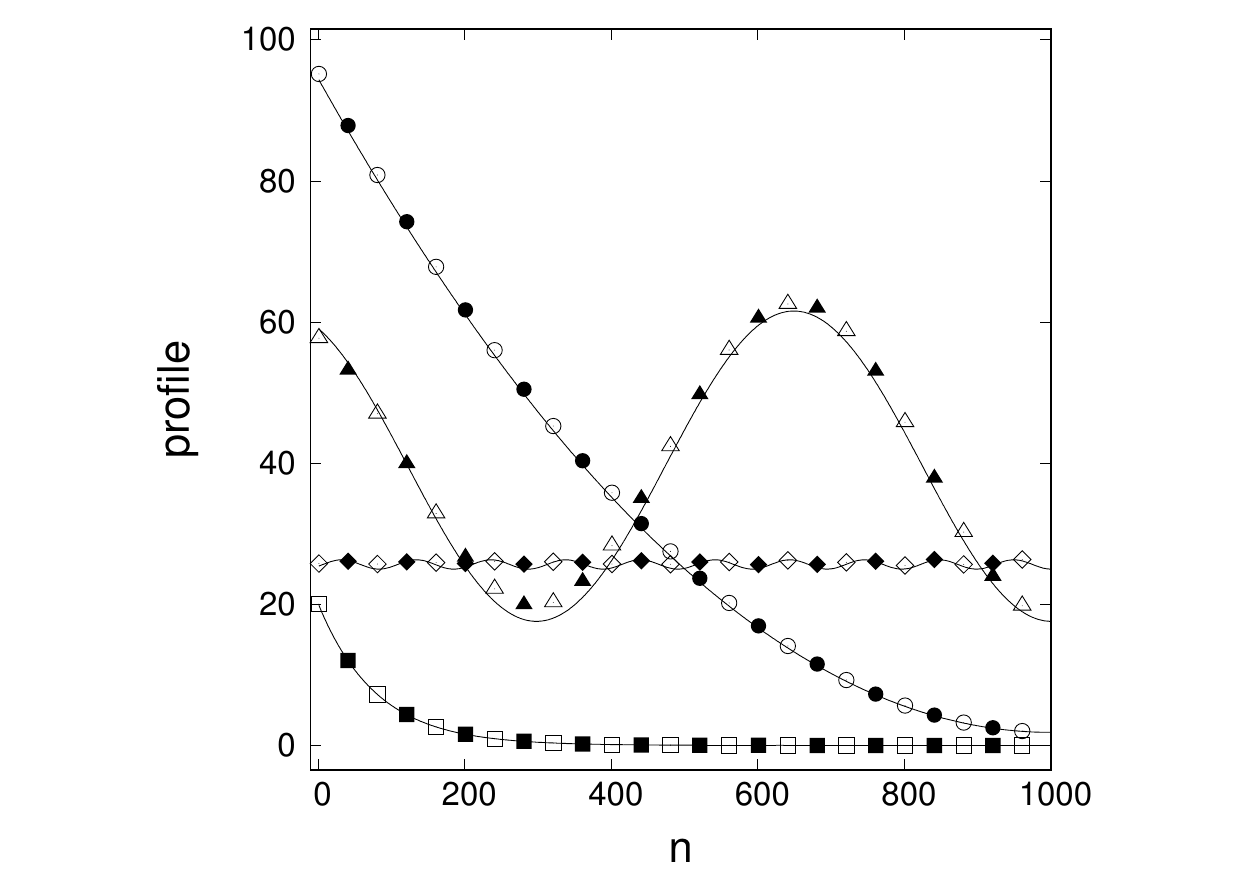} 
\caption{Profile of $\phi_n$ \eqref{qua110} for the regular Kronig--Penney model (solid line),  
numerical quenched average of $\phi_n$ for the
quantum disordered multi--barrier system 
(black symbols),
numerical quenched average of $\rho_n$ \eqref{ape070} for the ZRP model 
(open symbols).
The quenched average has been taken over $10^2$
random realizations of the multi--barrier system such that 
the parameters $p_n$ and $q_n$ are in the interval $(0,1)$, to let the ZRP model be well defined.
The values of the parameters are
$C=5$, $D=0$, $\gamma=1$, $L=10$, $V=1$, for energies
$E=0.1$ (squares), $E=0.5$ (circles), $E=0.7$ (triangles),
and $E=10$ (diamonds).
Shown are the cases with $N=50$ (left panel), $N=200$ (central panel) and $N=1000$ (right panel).
}
\label{fig:fig9}
\end{figure}

Lastly, we tested the connection between 
the regular quantum model and the quenched 
average of its disordered versions, constructed by drawing the random 
barrier widths under the constraints given in \eqref{constr1}.
To this aim, we considered a set of $10^2$
disordered realizations of the multi--barrier system with positive $p_n$ and $q_n$,
cf.\ \eqref{ape200}. In particular, we numerically computed both the quenched average of
$\phi_n$, coming from a straight implementation of the transfer matrix technique,
and the quenched average of $\rho_n$, obtained via MC simulations by averaging over
the stationary occupation profiles of the ZRP. Results for the quenched averages,
for different values of $E$ and $N$, are displayed in Fig.\ ~\ref{fig:fig9}. As expected, the
quenched average of $\rho_n$ perfectly agrees with the quenched average of $\phi_n$
for all the considered energies and numbers of barriers. We further observe that
the quenched disordered averages and the behavior of the regular Kronig-Penney models
also converge to each other for growing $N$. This reinforces the results of Refs.\ \cite{col1,col2,VanRon}, showing convergence with growing $N$ of the transmission coefficients of the random systems to the regular case values.

The correspondence highlighted in Figs.~\ref{fig:fig8} and \ref{fig:fig9}
 can be used to investigate, or to
interpret, some of the peculiar properties of the random Kronig--Penney model \cite{Drabkin12} with the
techniques developed in the study of the ZRP, and viceversa. In particular, the disordered
version of the quantum multi--barrier model, with the random variables $\lambda_n$, $\gamma_n$
scaled as in Eq.~\eqref{constr1}, can be studied as an instance of a random walk in a random environment
(with site randomness), for which a rich theory has been developed\cite{Solomon75,Sznitman03,Varadhan03,Zeitouni06}. On the
other hand, phenomena such as Sinai's localization \cite{Bogachev06}, which is observed in random walks
in random environments, can be investigated in terms of our random Kronig--Penney model.

\section{Conclusions}
\label{s:concl}
\par\noindent
In this paper we have shown that the quantum multi--barrier finite 
Kronig--Penney model and the ZRP are equivalent if the hopping probabilities of the stochastic 
process are properly tuned with the the parameters of the quantum system. 
In particular, we have evidenced that, for the finite Kronig--Penney model with unitary input only from the left boundary, 
the transmission 
coefficient corresponds to the stationary current in a boundary driven heterogeneous ZRP \cite{EH2005}, realized by 
independent particles performing a random walk on a lattice with 
site--dependent hopping probabilities.

If $p_n$ and $q_n$ are taken as prescribed by equation \eqref{ape200} and $\alpha$ and $\beta$ as prescribed
by equation \eqref{ape220}, the stationary profile $\phi_n$ 
and the current of the quantum model are correctly recovered from the ZRP, for any value of the energy $E$. 

To see how the choice of parameters is essential, let us drop the dependence of $p_n$ and $q_n$ on $S_n$
and $S_{n+1}$, in \eqref{ape200}. In this case, one may still recover the correct value of the 
stationary current but not the stationary profile for an arbitrary value of the energy.

It is reasonable to expect that our correspondence method produces the equivalence of the two models considered 
in this paper also in other situations. After all, everything in the ZRP depends on the hopping
probabilities and on the boundary conditions, like they depend on the boundary conditions and 
on the potential barriers in the quantum model. This may allow us to treat rigorously the 
large $N$ limit of mesoscopic disordered systems numerically investigated in 
Refs.~\cite{col1,col2,VanRon}, for which no general ergodic--like results 
seem to have been so far developed.

We have also shown that a random walk in a random environment
correctly reproduces the transport properties of the quantum disordered model. Thanks to this 
equivalence, one concludes for the stochastic process that the stationary profiles
may vary from monotonically decreasing to oscillating because of the variation of the
energy of an incoming plane wave, of the associated quantum multi--barrier model.
When this energy exceeds a critical value, the stationary states turn from monotonically
decaying to oscillating.

Finally, we also numerically observed the convergence, for $N$ large, of 
the quenched averages of the mean density $\phi_n$ and the stationary occupation $\rho_n$, for the disordered multi-barrier systems, 
to the corresponding value measured with the regular Kronig--Penney model.
This implies that
the quantum model has no localization for growing $N$. After all, 
this is natural in small systems, and reveals the different nature of our large $N$ limit, 
compared to the standard hydrodynamic limit.

The equivalence of quantum multi--barrier models and stochastic particle systems now 
suggests various avenues for future research, such as the investigation of 
\textit{uphill currents} in the quantum models. These kinds of currents, indeed, 
have been recently found in interacting particle (or spin) systems on lattices 
\cite{CDMP17,CDMP17bis,CC17,CGGV18,ACCG18}.

\vskip 20pt

\textbf{Acknowledgements.}
MC and ENMC acknowledge financial support from FFABR 2017.
LR acknowledges that the present research has been partially 
supported by MIUR grant Dipartimenti di Eccellenza 2018--2022.

\appendix
\renewcommand{\theequation}{\Alph{section}.\arabic{equation}}

\section{The $\gamma,L$--continuum limit of the hopping probabilities}
\label{s:appendix}
\par\noindent
In this appendix we consider the Kronig--Penney model 
introduced in the Section~\ref{s:mod} in the case $D=0$ (see the comment 
below \eqref{qua010}). 
As mentioned below the equation \eqref{zero}, see also \cite{CNP15},
the $\gamma,L$--\emph{continuum limit} is realized by keeping 
fixed all the parameters of the Kronig--Penney model but the 
number of barriers $N$, which tends to infinity.
Recalling the matrix $\mathbf{M}$ defined in \eqref{qua090}, 
following \cite{CNP15}, we let
\begin{displaymath}
\Phi = \Re{(M_{11})} = \cos(k\delta) \cosh(z\gamma\delta) 
+ \frac{z^2 - k^2}{2 k z} \sin(k\delta) \sinh(z\gamma\delta)
\;\;.
\end{displaymath}
Denoting the eigenvalues of $\mathbf{M}$ 
by $\mu_1$ and $\mu_2=\mu_1^{-1}$, 
one finds
\begin{displaymath} 
\mu_1=  \Phi - \sqrt{\Phi^2- 1} \quad \text{and} \quad
\mu_2=\Phi + \sqrt{\Phi^2- 1}
\end{displaymath}
which can be real or complex--valued, depending on the value of $\Phi$.
It is important to remark that $\mu_1\neq\mu_2$, indeed, by expanding 
$\Phi$ in Taylor series with respect to $\delta$ in a neighborood 
of $\delta=0$, one has 
\begin{displaymath}
\Phi(\delta)
=
1
-\frac{1}{2}L^2(E-E_0)\delta^2
+\frac{1}{24}L^4\Big[E^2-2EE_0+E_0^2\frac{\gamma(2+\gamma)}{(1+\gamma)^2}
                \Big]\delta^4
+o(\delta^4)
\;,
\end{displaymath}
which proves that, for $\delta$ small, 
$\Phi(\delta)<1$ (resp.\ $\Phi(\delta)>1$) 
for 
$E>E_0$ (resp.\ $E<E_0$).
Moreover, for $E=E_0$ the above expansion becomes 
$\Phi(\delta)=1-L^4E_0^2\delta^4/[24(1+\gamma)^2]+o(\delta^4)$ 
which proves that, for $\delta$ small, $\Phi(\delta)<1$ even for $E=E_0$.

In the sequel we shall often use the 
$n$--th power $\mathbf{M}^n$ of $\mathbf{M}$ for 
$n=1,\dots,N$. By slightly abusing the notation, 
we shall denote its elements by $M^N_{ij}$.
One can use the equation \eqref{qua080} to express 
the coefficients $C_{n+1}$ and $D_{n+1}$ in terms of the 
boundary condition $C$ (recall that we assumed $D_{N+1}=D=0$).
One first writes \eqref{qua080} for $n=N$ and 
finds $D_1=C M^N_{21}e^{-i2k\delta}/M^N_{11}$, then, using again 
\eqref{qua080} for a general $n$, one gets
\begin{equation}
\label{Cnp1}
C_{n+1}
=
C\frac{e^{- i k \ell_n}}{M^N_{11}} 
(M^N_{11}M^n_{22}-M^N_{21}M^n_{12})
\;\;\textrm{ and }\;\;
D_{n+1}
=
C\frac{e^{i k (\ell_n-2 \delta)}}{M^N_{11}} 
(M^N_{21}M^n_{11}-M^N_{11}M^n_{21})
\end{equation}
which
for $n=N$ reproduce $C_{N+1}$ and $D$, while for $n=0$
they yield $C$ and $D_{1}$. 

Recalling the definition of hopping probabilities
\eqref{ape200}, we wish to express in terms 
of the boundary condition $C$ the two quantities 
$|D_{n}|^2+S_{n}$ and $|C_{n+1}|^2+S_{n+1}$, respectively the average hopping rates 
to the left and to the right from the $n$--th site of the ZRP model. 
Using \eqref{Cnp1} we find 
\begin{displaymath}
\frac{|D_{n+1}|^2+S_{n+1}}{|C|^2}=
\frac{
|M^N_{21}M^n_{11}-M^N_{11}M^n_{21}|^2
+
\Re{[(M^N_{11}M^n_{22}-M^N_{21}M^n_{12})
(M^N_{12}M^n_{22}-M^N_{22}M^n_{12})
e^{ik\delta}
]}
}
{|M^N_{11}|^2}
\end{displaymath}
for $n=0,\dots,N-1$
and 
\begin{displaymath}
\frac{|C_{n+1}|^2+S_{n+1}}{|C|^2}
=
\frac{
|M^N_{11}M^n_{22}-M^N_{21}M^n_{12}|^2
+
\Re{[
(M^N_{11}M^n_{22}-M^N_{21}M^n_{12})
(M^N_{12}M^n_{22}-M^N_{22}M^n_{12}) 
e^{ik\delta}]}
}
{|M^N_{11}|^2}
\end{displaymath}
for $n=1,\dots,N$.

To compute the $N$ large limit of the quantities above express the 
$n$--th power of the matrix $\mathbf{M}$ as in \cite[Eq.~(3.15)]{CNP15}:
\begin{equation}
\label{app020}
\mathbf{M}^n
=
\frac{\mu_1^n-\mu_2^n}{\mu_1-\mu_2}\mathbf{M}
-
\frac{\mu_2\mu_1^n-\mu_1\mu_2^n}{\mu_1-\mu_2}\mathbf{I}
\,\,,
\end{equation}
where $\mathbf{I}$ is the identity matrix.

Recalling the definition \eqref{ape200} of right hopping probability $p_n$, 
we let $x=n/N\in(0,1]$ and, in the limit $N\to\infty$, we find  
\begin{equation}
p(x)=
\left\{
\begin{array}{ll}
\frac{1}{2}+\frac{E_0-E}{E_0(1+\cosh\left( 2 L\sqrt{E_0-E} (1-x)\right))-2 E}
&
E<E_0\\
\frac{1}{2}+\frac{1}{2+2 E_0 L^2 (1-x)^2}
&
E=E_0\\
\frac{1}{2}+\frac{E-E_0}{2 E -E_0(1+\cos\left( 2 L\sqrt{E-E_0} (1-x)\right))}
 &
E>E_0\\
\end{array}
\right.
\label{pgL2}
\end{equation}
One readily notes that $p(x) \in [0,1]$, as illustrated in Fig. \ref{fig:fig4}
(see, also, the comment at the end of the paragraph below \eqref{ape220}).

Finally, the $\gamma,L$--continuum limit of the stationary current 
\eqref{ape080}, 
takes the form:
\begin{equation}
\overline{J}=\lim_{N\rightarrow\infty}\left(|C_{n+1}|^2-(|D_{n+1}|^2 \right)=
\left\{
\begin{array}{ll}
\left| C\right|^2\frac{8 E (E-E_0)}{8 E (E-E_0)+E_0^2\left(1-\cosh\left( 2 L\sqrt{E_0-E} \right)\right)}&
E<E_0\\
\left| C\right|^2\frac{4}{4+E_0 L^2}
&
E=E_0 \\
\left| C\right|^2\frac{8 E (E-E_0)}{8 E (E-E_0)+E_0^2\left(1-\cos\left( 2 L\sqrt{E-E_0} \right)\right)} &
E>E_0\\
\end{array}
\right.
\label{Jbar}
\end{equation}
For $C=1$, the expression \eqref{Jbar} yields  
the asymptotic value of the transmission coefficient 
$\overline{S}$ obtained in \cite[Eq.~(4.3)]{CNP15}.

\bibliographystyle{spr-chicago}      
\bibliography{example}   

\end{document}